\documentclass[prd,aps,floats,floatfix,eqsecnum,nofootinbib,showkeys]{revtex4}
\usepackage{array,amsmath,amssymb,verbatim}
\usepackage{bm}
\usepackage{graphicx}
\usepackage{hyperref}
\usepackage{fontenc}

\newcommand{\avg}[1]{\langle #1 \rangle}

\newcommand{\bds}[1]{\boldsymbol{#1}}

\newcommand{\bx}{{\bds x}}

\newcommand{\br}{{\bds r}}

\newcommand{\bu}{{\bds u}}
\newcommand{\bv}{{\bds v}}
\newcommand{\lowsub}[2]{#1_{\raisebox{-1.2pt}{\scriptsize\rm #2}}}
\newcommand{\rhoM}{\lowsub{\rho}M}
\newcommand{\rhoB}{\lowsub{\rho}B}
\newcommand{\rhozB}{\lowsub{\rho}{0B}}

\begin{document}
\title{Hollow cores in Warm Dark Matter halos from the Vlasov--Poisson equation}
\author{\bf Claudio Destri}
\affiliation{Dipartimento di Fisica G. Occhialini, Universit\`a
Milano-Bicocca\\ and INFN, sezione di Milano-Bicocca, Piazza della Scienza 3,
20126 Milano, Italia.\\
e-mail: claudio.destri@mib.infn.it}
\date{\today}
\begin{abstract}
  We report the results of extended high--resolution numerical integrations of the
  Vlasov--Poisson equation for the collapse of spherically symmetric WDM halos. For
  thermal relics with mass $m=1\,$keV/$c^2$, we find collapsed halos with cores of size
  from $0.1$ to $0.6\,$kpc. The typical core is hollow, with the mass density decreasing
  towards the core center by almost three orders of magnitude from its maximum near the
  core radius $r_{\rm c}$. The core is in equilibrium with the diffused part of the halo
  but far from virialization. These properties are rooted in the conservation of the
  squared angular momentum and in the original excess, proper of WDM initial conditions,
  of kinetic energy in the core region. In a sample of more than one hundred simulated
  collapses, the values of $r_{\rm c}$ and of the core density $\rho_{\rm c}$ are in the
  range typical of dwarf spheroids, while the maximal circular velocities $V_{\rm max}$
  are proper of small disk galaxies. The product $\mu_{\rm c}=\rho_{\rm c}r_{\rm c}$ takes
  values between $116\,M_\odot/$pc$^2$ and $283\,M_\odot/$pc$^2$, while the surface
  density $\mu_0$, as determined from a Burkert fit, is roughly three times larger. From
  these data and data obtained at smaller values of $m$, we extrapolate for one particular
  halo $\mu_{\rm c}=263(308)\,M_\odot/$pc$^2$ and $\mu_0=754(855)\,M_\odot/$pc$^2$ at
  $m=2(3.3)\,$keV/$c^2$, to be compared with the observed value $140^{+83}_{-52}
  M_\odot$/pc$^2$. In view of the many improvements and enhancements available, we
  conclude that WDM is a viable solution for explaining the presence and the size of cores
  in low mass galaxies.
\end{abstract}
\maketitle
\tableofcontents

\section{Introduction, summary and outlook}

The investigation on how purely self--gravitating matter evolves on long time scales has
played and still plays a central role in cosmology, in astrophysics and in statistical
mechanics. From the statistical point of view, the main obstacle towards the establishment
of a general picture is the lack of simple additivity due to the long range interaction.
This prevents a system, even if composed by a macroscopic number of ``particles'', to
reach thermodynamic equilibrium, so that the usual laws of equilibrium thermodynamics do
not apply \cite{lbell2}. Conversely, it is well known that there exist infinitely many
stable equilibrium phase--space configurations to choose from as $t\to\infty$ \cite{bt}.

Still, relaxation of some sort must occur, since self--gravitating systems of many
different kinds and sizes appear to be in (quasi--) stationary states
\cite{padma,bt}. Likewise, numerical $N-$body simulations of such systems show
that, from a coarse--grained point of view, some type of equilibration does occur
\cite{aars}. The central issue is then how to properly identify the quasi--stationary
states corresponding to given initial conditions. As detailed below, such identification
is the aim of this work in the specific case of single Warm Dark Matter (WDM) halos,
although the implications of our findings probably extend to wider domains.

Indeed, in the cosmological context, important questions concern the properties of DM
halos, which are the cradles of galaxy formation \cite{wr}. In turn, the halos are formed
through gravitational collapse triggered by tiny perturbations over the uniform DM
background which dominates the total matter contribution $\Omega_{\rm M} \sim 0.3$ to the
energy density of the Universe (baryons contributing only a smaller fraction $\Omega_{\rm
  b}\sim 0.04$). In this scenario, the counting and the properties of galaxies with any
given mass should be traced back to the primordial power spectrum of DM perturbations,
which statistically determines the initial conditions, and to the quasi--stationary state
(or states) to which gravity drives the various local matter overdensities.

For example, in a Universe dominated by Cold Dark Matter (CDM) the density perturbations
are gravitationally unstable down to mass scales much smaller than those of galaxies and
extensive numerical simulations (see {\em e.g.} \cite{fw} and references therein) have
shown that structure formation proceeds bottom--up with the early collapse of very small
regions, followed by larger and larger ones, with a complex merging histories that produce
successively more or less relaxed objects of increasing mass, like dwarf satellites,
massive galaxies, groups and clusters. Thus the properties of a single DM halo depend on
the full history of a much larger region of comoving space, hindering a clear--cut analytic
modeling of its quasi--stationary state.
 
However, this CDM scenario has (at least) two distinguishing features contradicted by
observations: it exhibits an overabundance of small mass virialized halos
\cite{sat1,sat2,sat3,sat4,sat5} and cuspy density profiles \cite{nfw,moore} where
observations point to cored ones \cite{obs1,obs2,obs3,obs4,obs5,obs6}. The latter problem,
the so--called cusp--core puzzle of CDM, apparently fades away upon including baryon
feedback in certain simulations \cite{bbr1,bbr2,bbr3}, but not in others \cite{bbr4,bbr5}.

\medskip

Structure formation based on WDM is in principle protected against these CDM difficulties,
while reproducing all the observed large scale ($\gtrsim 10\,$Mpc) properties as CDM does.
In fact WDM is characterized by the proper amount of primordial velocity dispersion, with
associated free streaming, to smooth out initial density fluctuations over scales below
the Mpc \cite{cdw,bode}. This velocity dispersion should also help to better match the
observed density profile of collapsed halos. In brief, WDM reconciles, or tend to reconcile
with observations several small to middle scale astrophysical and cosmological features
such as, for example, the halo concentration and satellite counts in $N-$body simulations
\cite{bode,avila}, the phase--space density of dwarf spheroids in a model--independent WDM
treatment \cite{bdvs,dvs}, the mass function in the halo model \cite{smark}, the galaxy
luminosity functions and the stellar mass distributions in semi--analytic models
\cite{menci}. Many other investigations \cite{inv1,inv2,inv3,inv4,inv5,inv6,inv7} confirm
that WDM agrees with observations better than CDM.

Still, one observational fact that many numerical simulations on WDM fail to correctly
reproduce is the size of the halo cores
\cite{smco1,smco2,smco3,smco4,smco5,smco6,smco7}. Indeed, a well known theoretical argument,
rooted on the Tremaine--Gunn bound \cite{trgb}, implies  that WDM halos are cored
\cite{hodal,smco5}. It is based on the so--called (pseudo) phase--space density
\cite{psd1,psd2,hodal,psd4,psd5}, $Q = \rho/\sigma^3$, where $\rho$ is the characteristic
mass density and $\sigma$ the characteristic one--dimensional velocity dispersion of the
core. One can argue \cite{hodal} (see \cite{bri,smco6} for more punctual analysis that
however do not alter significantly the conclusions) that $Q<Q_{\rm prim}$, where the
space--constant $Q_{\rm prim}$ is computed at WDM decoupling, in any given particle
model. Then, {\em assuming} a halo model ({\em e.g} an isothermal sphere as in
ref.~\cite{hodal} or a pseudo-isothermal one as in ref.~\cite{smco5} to relate $\sigma$ to
$\rho$, one obtains a bound on the core size. WDM $N-$body simulations apparently produce
cores which almost saturate this bound \cite{smco5,smco6}. The problem is that these cores
are resolved, and found with size comparable to the observed ones, only for initial velocity
dispersions so large that the corresponding free streaming would have erased at the linear
level the fluctuation seeds of the halo themselves (the {\em catch 22} of
ref.~\cite{smco5}). Viceversa, using the theoretical bound to extrapolate the size of
resolved cores to velocity dispersions small enough to allow the fluctuation seeds, yields
cores two order of magnitude smaller than the observed ones.

If this were indeed the situation, WDM structure formation would need help from baryon
feedback (if that really can help) almost as badly as CDM does. Alternatively, in could be
that current $N-$body simulations, plus extrapolations based on equilibrium assumptions,
do not provide a quantitatively sound description of WDM gravitational collapse below the
kpc scale. Here we address the second possibility only and give an affirmative answer, by
showing that WDM is an effective working hypothesis for explaining the size of observed DM
cores.

\medskip 

In this work, we report the results of extensive high--resolution direct numerical
integrations of the spherically symmetric Vlasov--Poisson equation for a $1\,$keV WDM thermal
relic. The Vlasov--Poisson equation (\ref{eq:VP}), often called collisionless
Boltzmann--Poisson equation, describes the non--dissipative incompressible $6-$dimensional
phase--space flow of self--gravitating continuous matter. It is what $N-$body simulations
attempt to approximately solve with fictitious particles moving in $3-$dimensional
configuration space.

Our findings, outlined in the next subsection and described in more detail in
Sec.~\ref{sec:results}, provide an overall picture of WDM halo cores quite different from
the commonly expected one, which is based on some apriori assumptions and the (extrapolation of
the) results of $N-$body simulations.

\subsection{Non--virialized hollow cores with nearly constant surface density}
\label{sec:nonvir}

A common expectation about cored DM halos, as those of WDM, is that their mass density is a
monotonically decreasing function of $r$ which has some core radius $r_{\rm c}$ as unique
length scale in the core region. This is indeed the situation in collisionless
self--gravitating systems at equilibrium, that are described by stationary and stable
ergodic phase--space distributions \cite{bt}. The above expectation is then
extended to quasi--stationary states, with the understanding that they do not differ much
from truly equilibrium states. The property that should guarantee this proximity is core 
virialization, which can be described as follows. 

Let us consider a spherical, purely self--gravitating system. Thanks to Gauss' law, one
can unambiguously compute the gravitational potential energy $U(r)$ within the sphere of
radius $r$, regardless of the system configuration for $r'>r$. Hence one can define the
$r-$dependent virial ratio $W(r)=-2K(r)/U(r)$, where $K(r)$ is the kinetic energy within
the sphere. Suppose the system has relaxed to some quasi--stationary state with a core of
size $r_{\rm c}$.  For $r$ large enough we expect virialization, that is
$W(r)\gtrsim1$. For $r\ll r_{\rm c}$ instead, we have $W(r)\sim r^{2}$, since $K(r)\sim
r^{-3}$ while $U(r)\sim r^{-5}$ as $r\to0$ in a core that is in hydrostatic
equilibrium. In between $W(r)$ will decrease in some system--specific way.  If
$W(r)\gtrsim 1$ for $r/r_{\rm c} \gtrsim 1$, we can say that also the core is
virialized. Since $r_{\rm c}$ is the only scale in the core region, also $W(r)$ is a
function only of $x=r/r_{\rm c}$, which behaves as $x^{-2}$ for $x$ small enough and
monotonically decreases in the interval $0<x\lesssim1$.  This is exactly what happens for
instance in the isothermal sphere and many other systems with ergodic phase--space
distributions.

CDM halos do not have cores but are quite virialized, so they have only the
system--specific decrease. For example, if we rather crudely assume $\rho\sim r^{-1}$ {\em
  \`a la} NFW \cite{nfw} and $Q\sim r^{-1.9}$ \cite{hoff}, then $W(r)\sim r^{-0.4}$, in
any case much slower than within a core.

WDM halos are cored and the $r< r_{\rm c}$ region can be theoretically and/or numerically
investigated, provided the resolution is high enough.  According to the common lore just
described, the inner part of a halo should be in some quasi--stationary state not too different
from stable equilibrium states. The halo core would then be virialized in the
sense defined above.

\medskip

On the contrary, the main result of our investigation can be stated as follow: {\em The
  spherical collapse of WDM yields cores that are hollow and not virialized}.

Details are provided in Sec.~\ref{sec:results} (see Figs.~\ref{fig:sixp}, \ref{fig:rho}
and \ref{fig:W} in particular). Here we just make few simple observations to
support these numerical findings.

\medskip

Let us assume for the moment that the inner part of the halo can indeed be modeled by
some equilibrium--type cored configuration. Then for $r\lesssim r_{\rm c}$ we have
$W(r)\lesssim (r_{\rm c}/r)^2$, where $r_{\rm c}$ is in the kpc scale to match, for
instance, the DM halos of dwarf galaxies.

Consider now the uniform isotropic Universe at some time $t=t_0$ when the energy
contribution of matter inhomogeneities is negligible and choose, in order to
evaluate the initial value $W_0(r)$ of the virial ratio, the center of a future collapse
(of what type of DM does not matter yet) as origin of the comoving coordinates. The result
is (see Sec.~\ref{sec:W} for details on the elementary derivation):
\begin{equation*}
  W_0(r) = 2 + \frac{10\sigma_0^2}{a^4H^2r^2} \;,
\end{equation*}
where $a$ is the scale factor, $H={\dot a}/a$ is the Hubble parameter and $\sigma_0$ is the
DM (comoving) velocity dispersion. In the derivation we assumed that DM is
non--relativistic at the time $t_0$ considered.

In the case of CDM, we have $\sigma_0=0$ and the gravitational collapse will need to
reduce the nearly constant $W_0$ by just a factor $2$ to virialize the halo. Instead, in
the case of WDM particles, that decouple while ultra--relativistic but are already
non--relativistic deep in the RD era, we obtain
\begin{equation*}
  W_0(r) \simeq 2 + 10\left(\frac{l_{\rm fs}}{r}\right)^{\!\!2}\quad,
  \qquad l_{\rm fs} = \left(\Omega_{\rm r}\right)^{-1/2}\frac{\sigma_0}{H_0} \;,
\end{equation*}
where we used $a^4H^2 \simeq\Omega_{\rm R}H_0^2$, with $\Omega_{\rm R}$ the radiation
fraction at time $t=t_0$, to recognize in $l_{\rm fs}$ a (very crude) estimation of the
free--streaming length of the WDM particles. Therefore
\begin{equation*}
  \frac{W_0(r)}{W(r)} \simeq 10 \left(\frac{l_{\rm fs}}{r_{\rm c}}\right)^{\!\!2}
  \quad,\qquad r \lesssim r_{\rm c} \;.
\end{equation*}
Indeed, by construction $r_{\rm c}\ll l_{\rm fs}$. For instance, thermal relics with a
mass of $1\,$kev/$c^2$ (and two internal degrees of freedom) have $\sigma_0= 0.025~$km/s,
so that $l_{\rm fs}\simeq 40~$kpc since $\Omega_{\rm R}=8.5\times 10^{-5}$ by the time
only neutrinos and photons contribute to radiation. Hence, throughout the region of the
future core, kinetic energy initially exceeds by three orders of magnitudes the typical
value it would eventually have in a few kpc virialized core. And the smaller the core, the
higher the kinetic energy excess, since $l_{\rm fs}$ is fixed way before the collapse
started.

Of course this is an idealized scenario, since the real collapse will not be spherical,
there could be mergers and so on. But the free streaming of a keV-sized WDM has also the
merit of smoothing out the matter fluctuations on scales of several hundreds kpc. Then the
possibility of a nearly spherical and almost undisturbed collapse is certainly not a
far--fetched idealization as it would be for CDM. Hence the kinetic energy
excess in the initial conditions w.r.t. the virialized core after the collapse holds
beyond the approximation of spherical symmetry.

During the collapse, phase mixing and violent relaxation \cite{vrelax1} alone might still
bring the system to a quasi--stationary state but, lacking more efficient energy transfer
mechanisms such as radiation or dissipation, they need not be able to reduce the virial
ratio by the thousands in the finite amount of time available. This is just what we
observe by numerically integrating the Vlasov--Poisson equation.

A complementary viewpoint is based on the conservation of angular momentum, which is the
same in physical and comoving coordinates. In a spherically symmetric system, the
$1-$particle angular momentum is conserved also if the potential is not constant.  Of
course, the mean vector angular momentum vanishes by symmetry, but the squared angular
momentum $\ell^2$ does not. Matter which is initially at a distance $R\gg r_{\rm c}$ but
will eventually fall in the core, has a conserved $\ell^2$ of order $(R\sigma_0)^2$. Like
the skate dancer that closes her arms, this matter will spin faster in the core,
sustaining the original excess of core kinetic energy.

\medskip

The net result we observe and describe in Sec.~\ref{sec:results}, is the formation of
halos with a hollow core, that is with a non monotonically decreasing mass density which
develops its maximum $\rho_{\rm max}$ at some $r_{\rm max}>0$, which provides a first
natural definition of core radius. Throughout the core and beyond a large excess of
kinetic energy is trapped mostly in tangential motions. The core is approaching a dynamic
rather than hydrostatic equilibrium and the mass density $\rho(r)$ features a roughly
parabolic rise till $r_{\rm max}$, where $\rho_{\rm max}$ is almost three order of
magnitude larger than the density deep into the core. For $r>r_{\rm max}$, $\rho(r)$ quite
rapidly settles to the $r^{-2}$ decrease that implies a constant circular
velocity. Clearly the crossover at $r=r_{\rm max}$ provides another length
scale beside $r_{\rm max}$.

We find that these properties are common to all the 111 halos of our sample. The halos
are obtained starting from different initial overdensity profiles, which are in turn
constructed, according to the procedure described in Sec.~\ref{sec:sphinit}, by angle
averaging over the peaks of a random realization of the fluctuation field proper of a
$1\,$keV WDM thermal relic. The overall shape of $\rho(r)$ is almost independent of the
initial conditions and of the location or value of the maximum itself (see
Figs.~\ref{fig:sixp}, \ref{fig:rho} and \ref{fig:highres}). These latter quantities
instead, which fix the overall scale of $\rho(r)$, do depend on the details of the the
initial fluctuation profile, such as height and size.

The hollow core shape allows a natural definition of the core radius $r_{\rm c}$ and the
core density $\rho_{\rm c}$ slightly different from $(r_{\rm max},\rho_{\rm max})$: the
pair $(r_{\rm c},\rho_{\rm c})$ define the point where the density $\rho(r)$ should be cut
to replace the hollow core with a constant density core (see Sec.~\ref{sec:rc}).  $r_{\rm
  c}$ is therefore slightly larger than $r_{\rm max}$.

\medskip

In our sample we find values of $r_{\rm c}$ ranging from $0.1$ to $0.6\,$kpc and values of
$\rho_{\rm c}$ ranging from $0.2$ to $2.5\,M_\odot/$pc$^3$. Very interestingly, the core
radius $r_{\rm c}$ is nearly inversely proportional to the core density $\rho_{\rm c}$
leading to an nearly constant value of $\mu_{\rm c}\equiv\rho_{\rm c}r_{\rm c}$ around
$210\, M_\odot/$pc$^2$ (see Sec.~\ref{sec:rc} for more details), in remarkable agreement
with the observations reported in refs.~\cite{surd1,surd2,surd3,surd4}.  However, while
the near constancy of $\mu_{\rm c}$ could be rooted in the core hollowness (a constant
surface density means a constant core mass per unit area, a property certainly more
appropriate to hollow cores than to bulky ones), the quantitative proximity to the
observational value $\mu_{\rm 0,obs}=140^{+83}_{-52} M_\odot$/pc$^2$ is rather puzzling
(and quite intriguing), since the estimation of $\mu_{\rm 0,obs}$ is based on the
hypothesis that the DM density has a bulky Burkert profile \cite{burk}, with its own
definition of core radius $r_{\rm 0B}$ and core density $\rho_{\rm 0B}$.  When a Burkert
fit is performed on the hollow cores for $r_{\rm c}\lesssim r\lesssim 10\,r_{\rm c}$ (see
Sec.~\ref{sec:rc} for the details), one obtains values of $\mu_0\equiv\rho_{\rm 0B}r_{\rm
  0B}$ around $600\ M_\odot/$pc$^2$, roughly four times larger than $\mu_{\rm 0,obs}$.
\medskip

The typical $r_{\rm c}$ we find is more than one order of magnitude larger than what
expected for a WDM particle of mass $m=1\,$keV/$c^2$ according to the $Q-$based
theoretical bound and extrapolations from resolved cores of $N-$body simulations
\cite{hodal,smco5,smco6}. But this type of arguments rely on the apriori assumption of a
{\em thermalized}, and {\em a fortiori} virialized, core and therefore is not applicable
to the WDM cores found in our simulations. See Sec.~\ref{sec:rc} for the explicit
comparison of the hollow core with that of the isothermal sphere, which shows in detail
how the isothermal extrapolation fails; see Sec.~\ref{sec:sigma} for our technical
explanation of why it fails.

On the other hand, at odd with $N-$body simulations, our halos feature a diffuse part with
a $r^{-2}$ tail which is too long. The density is still decreasing almost as slowly as $r^{-2}$
where it tales values from 100 to 200 times larger than the critical density. One possible
cause is a bias in our construction of spherically symmetric overdensity profiles
for the initial conditions (see Sec.~\ref{sec:sphinit}), which does not allow enough
matter outflow from the collapsing halo. This makes it awkward to build a reliable $r_{\rm
  c}$ {\em vs.}  $M_{\rm halo}$ relation, for any common definition of halo mass $M_{\rm
  halo}$. 

Hence, as in observations, the best way to quantify the mass content of a halo is to
consider the maximal circular velocity $V_{\rm max}$. In our sample we find that $V_{\rm
  max}$ takes values between $50$ to $70\,$km/s, characteristic of small disk galaxies,
whereas $r_{\rm c}$ and $\rho_{\rm c}$ are typical of dwarf spheroids. This concentration
excess is properly measured by means of the Burkert fit for $r_{\rm c}\lesssim r\lesssim
10\,r_{\rm c}$. We find that $r_{\rm cB}$ is roughly twice $r_{\rm c}$, while $V_{\rm
  B,max}$, the maximal circular velocity of the Burkert fit, is always very close to
$V_{\rm max}$. Hence, apart from the mass deficit of the hollow core w.r.t. the Burkert
core for $r<0.5\,r_{\rm cB}$, our halos differ from realistic DM halo only because the
surface density $\mu_0$ is four times larger than the observed value. In other words, the
region $0.5\,r_{\rm 0B}\lesssim r\lesssim 5\,r_{\rm 0B}$ of our hollow--core halos is more
concentrated than real DM halos by a factor roughly equal to $4$. In Sec.~\ref{sec:sigma}
we estimate that this factor grows to $5$ when the mass of the WDM particle is raised up
to $m=2~$keV$/c^2$, the value that now seems most favored \cite{dvs2,abaz}, and reaches
$6$ when $m=3.3~$keV$/c^2$, the lower bound from Lyman--$\alpha$ and hydrodynamical
simulations of ref.~\cite{viel}.

This quantitative analysis is the least restrictive, since the Burkert fit is performed
only for $0.5\,r_{\rm 0B} \lesssim r \lesssim 5\,r_{\rm 0B}$, the only region of the inner
halo where the fit can be accurate. Since the bulky Burkert profile provides a very good
fit to real DM halos down to few percents of $r_{\rm 0B}$, the hollow core is most likely
ruled out observations. Hence whatever improvement, modification and/or enhancement that
might succeed in reducing $\mu_0$ should also eliminate, or strongly reduce, the
hollowness. An obvious addition could be baryons, completely neglected in the zeroth
order approach of this work and possible hollowness--reducer thorough adiabatic
compression. Another could be quantum corrections, to be briefly motivated in the next
section.

\medskip

At any rate, since we find cores with an inner mass deficit rather than excess, with a
surface density nearly constant and relatively close to the observed value, the
starting point provided by our VP approach appears by far better than in CDM--only
$N-$body simulations or in the too-small--core WDM simulations cited above.

In the next section we outline some improvements for our WDM-only VP simulations and we
also question the soundness of another common apriori assumption on DM dynamics.

\subsection{Directions for improvement}\label{sec:improve}

The first, obvious and major improvement would be to go beyond the approximation of
spherically symmetric collapse and tackle the full $6-$dimensional Vlasov--Poisson
problem. A promising low--resolution attempt in this direction is reported in
ref.~\cite{yoshi}, but the computational resources required for a high--resolution
cosmological simulation appear at the moment prohibitively large, at least for the author.
Moreover, it is very unlikely that the problem $\mu_0/\mu_{\rm 0,obs}\sim 4$ is entirely
due to the approximation of spherical symmetry.

Remaining in the more tractable framework of spherical symmetry, we can consider a few
interesting directions for future improvement:
\begin{enumerate}
\item A problematic aspect of our halo sample is the small variation of the basic
  parameters $\rho_{\rm c}$, $r_{\rm c}$ and $V_{\rm max}$. The most likely cause is a
  bias in our construction of spherically symmetric overdensity profiles for the initial
  conditions (see Sec.~\ref{sec:sphinit}), which does not allow enough matter outflow from
  the collapsing halo. A more sophisticated selection procedure could be implemented to
  better mimic, even within the approximation of spherical symmetry, the effects of nearby
  collapsing halos on the halo of interest. This could at the same time increase the
  parameter variations and reduce the too long $r^{-2}$ tail of the halos. It is also
  conceivable that a shorter $r^{-2}$ tail could improve the value of the surface density
  $\mu_0$, although probably less than necessary.

\item The large initial virial ratio and the crucial role played by angular momentum imply
  that the halo collapse might be sensitive to details of the freezed--out velocity
  distribution, which in our simulation we fixed to the simplest Fermi--Dirac form proper
  for thermal relics (see Sec.~\ref{sec:VPDM}).  Indeed we find that the core properties
  strongly depend on $\sigma_0$ (see Sec.~\ref{sec:sigma}. Hence other WDM models, which
  are already known to yield different cutoff effects on the primordial power spectrum
  \cite{dan,petra,danw,dvs3}, might also lead to noticeable effects on the core
  properties. In this respect, accurate analysis at the linear level also of WDM
  velocities, as that in ref.~\cite{dan2}, could play an important quantitative role in
  the collapse dynamics through a better determination of the initial conditions.
\item Another common apriori assumption on DM dynamics, besides the core virialization
  discussed above, is that quantum effects are fully negligible. The large density
  variation in the hollow core, the crucial role played by angular momentum in our WDM
  collapse and the profound quantum modifications to the theory of angular momentum,
  suggest instead that Quantum Mechanics could have a deep role in the shaping of WDM
  cores, provided its effects were non negligible in the primordial Universe. We elaborate
  this point below. The relevance of Quantum Mechanics for ultra--compact dwarf galaxies
  with WDM halos, within an equilibrium approach, was discussed already in ref.~\cite{ddvs}.
\end{enumerate}

\medskip

For a particle of mass $m$, the quantum unit of phase--space density is
\begin{equation*}
  q = \frac{m^4}{(2\pi\hbar)^3} = 5.13 \times 10^{-4} \left(\frac{mc^2}{\rm keV}\right)^4
  \frac{M_\odot}{{\rm pc}^3} \,({\rm km}/{\rm s})^{-3} \;. 
\end{equation*}
Strictly speaking it should just be $(2\pi\hbar)^{-3}$, the inverse of the volume of a
quantum cell in the phase space with the conventional dimensions of (length $\times$
momentum)$^3$. The extra factor of $m^4$ comes from the normalization to a mass density,
rather than a number density, and to the use of velocity rather than momentum as coordinate
in phase space. If $g$ denotes the number of non--translation degrees of freedom of the
particle, such as spin and/or internal symmetry quantum numbers, it follows that (notice
the $(2\pi)^3/g$ difference w.r.t. ref.~\cite{ddvs})
\begin{equation*}
  \frac{Q}{g\,q} = \frac{(2\pi\hbar)^3\rho}{g\,m^4\sigma^3} = 
  \frac1{g}\left(\frac{\lambda_{\rm dB}}{d}\right)^{\!3} \;,\quad \lambda_{\rm dB} =
  \frac{2\pi\hbar}{m\sigma}\;,\quad d = \left(\frac{m}\rho\right)^{\!1/3}
\end{equation*}
where $\lambda_{\rm dB}$ is the characteristic de Broglie wavelength and $d$ is the mean
interparticle distance. Thus $Q/(gq)$ provides a semiclassical measure of how much
particles in a non--relativistic gas with density $\rho$ and velocity dispersion $\sigma$
are packed w.r.t. the reference quantum packing fixed by Heisenberg's indetermination
principle. Values $\gtrsim 1$ of $Q/(gq)$ then indicate that quantum effects could be
important, as in the standard non--relativistic free Fermi gas.

Since $Q$ largely decreases during the classical gravitational collapse (we verified this
also in our simulations, see Sec.~\ref{sec:phiQ}), we have to compare it with $gq$ at the
beginning, when WDM is already non--relativistic but still nearly homogeneous. For thermal
relics the issue is solved in the simplest possible way, since in that case the primordial
value of $Q/(gq)$ is a pure number that does not depend on cosmological parameters nor on
the actual value of the mass: \cite{hodal,bdvs,dvs}:
\begin{equation*}
  \frac{Q_{\rm prim}}{g\,q} = 4\pi\sqrt{27}\; \frac{I_2^{5/2}}{I_4^{3/2}} = 2.52950728\ldots
  \;,\qquad I_n \equiv \int_0^{\infty}\frac{x^ndx}{1+e^x} \;.
\end{equation*}
Thus the classical framework might not be fully adequate for WDM, after all. If so, also
the history before should be revisited to determine more appropriate initial conditions
with the necessary quantum corrections, Namely, the classical framework ignores from the
start the fermionic quantum pressure which, in view of the above value of $Q_{\rm prim}$,
could significantly contrast the gravitational pull and flatten the halo cores, as
advocated in refs.~\cite{ddvs,bidvs}.

\section{Vlasov--Poisson equation}

In the mean--field approximation, the phase--space one--particle distribution function
$f(\br,\bv,t)$ of a purely self--gravitating system of $N$ identical particles evolves
according to the Vlasov--Poisson (VP) equation
\begin{equation}\label{eq:VP}
  \begin{split}
    &\big[ \partial_t + \bv\cdot\nabla_r + (\nabla_r\Phi)\cdot\nabla_v \big] f = 0\;,\\
    &\nabla_r^2\Phi = 4\pi G\,\rho \;,\quad \rho(\br,t) = \int
    d^3v\,f(\br,\bv,t)\;.
  \end{split}
\end{equation} 
Here $\rho$ represent the mass density of the system, obviously normalized as 
\begin{equation*}
  \int d^3r\, \rho(\br,t) = N m \equiv M \;,
\end{equation*}
where $ m$ is the particle mass. Strictly speaking then, it is $f/m$ that plays the role
of one--particle distribution function. Mean--field approximation means that two--body or
higher correlation functions are considered always negligible. Then the specific value $m$
of the particle mass drops completely out of the game if $f$ is normalized as in
eqs.~(\ref{eq:VP}), which now describe the evolution of a non--dissipative incompressible
fluid in phase space subject only to the self--consistent gravitational force it
generates. In fact, the first equation in (\ref{eq:VP}) is nothing but Liouville's
equation for free streaming in the ``external'' field $\Phi$.
 
It is commonly believed (and in a certain special sense rigorously demonstrated
\cite{braun}) that in the limit $N\to\infty$ the mean--field description becomes
exact. But from the more practical point of view of understanding the dynamics of the
$10^{70}$ (or more) particles of a DM halo, for example, there can be little doubt that the
self--gravitating fluid description implied by the mean--field approximation is more
accurate than any $N'-$body dynamics with $N'$ at least $10^{60}$ times smaller than $N$
and simulation ``particles'' as massive as $10^4 M_\odot\sim 10^{50}$ GeV/$c^2$. Due to
the absence of dissipative phenomena in the collisionless mean--field dynamics, it is
legitimate to have doubts also on the hydrodynamic approximations to
eqs.~(\ref{eq:VP}), where the moment expansion in velocity space is closed at second
order by some heuristic equation of state.
  
The problem, from a numerical point of view, is that simulations of the six--dimensional
continuous system eqs.~(\ref{eq:VP}), without any symmetry to help, are much
more demanding than $N'-$body or $3-$dimensional hydrodynamical simulations.

Since the Vlasov equation is formally time--reversible, the system retains full memory of
its initial conditions, albeit dispersed on smaller and smaller scales over phase
space. It is commonly believed that any reasonably coarse--grained version of the exact
solution $f(\br,\bv,t)$ does relax to some form of equilibrium, or quasi--stationary
states (QSS), a typical feature of systems with long-range interactions. Lynden--Bell
theory of violent relaxation \cite{vrelax1,vrelax2,vrelax3} provides a rather general
framework for this, although only for very special situations it is has been possible to
verify that the QSS to which the system relaxes does indeed maximize the Lynden--Bell
coarse--grained entropy \cite{levin,sylos}.

In other words, little is known on the class of initial conditions that do lead to QSS,
or, if the system does relax to QSS, on the map from the space of initial conditions to
the space of QSS and on the relative time scales. However, this fact is not a major issue
in the cosmological applications of the VP equation, since the initial conditions in the
linear regime of the gravitational clustering are rather well known. Then the real problem
is to solve as accurately as possible the VP equation and see what really happens in the
finite cosmic time available, regardless of any assumed relaxation to a QSS.

\subsection{VP equation for Dark Matter}\label{sec:VPDM}

In the cosmological FRW spacetime, well after matter-radiation equilibration, the
non--relativistic DM fluid evolves according to VP system (\ref{eq:VP}) of Newtonian
equations, with $t$ identified with the cosmic time. But in a cosmological context one
cannot ignore the (accelerated) universe expansion even for a single DM halo.  If the
phase--space coordinates $(\br,\bv)$ are interpreted as physical coordinates for a DM
particle, then no change is needed in eqs.~(\ref{eq:VP}) to implement a decelerated
expansion, since the latter affects only the initial conditions through the assumption of
the proper Hubble flow.  On the other hand, the accelerated expansion caused by the
cosmological constant $\Lambda$ requires to introduce antigravity, by appropriately
changing the source term in Poisson's equation, that is
\begin{equation}\label{eq:poi2}
  \nabla_r^2\Phi = 4\pi G (\rho - 2\rho_\Lambda) \;, \quad 
\end{equation}  
where 
\begin{equation*}
  \rho_\Lambda = \Omega_\Lambda\rho_{\rm crit} \;, \quad \rho_{\rm crit} = 
  \frac{3\, H_0^2}{8\pi G}
\end{equation*}
is the energy density due to $\Lambda$, written in terms of the present fraction
$\Omega_\Lambda$ and of the critical density or, equivalently, of the present Hubble
parameter $H_0$. Here we neglect the other smaller source of Newtonian potential, namely
baryons, so that $\Omega_{\rm DM} = \Omega_{\rm M}$ and $\Omega_\Lambda = 1-\Omega_{\rm
  M}\simeq 0.7$ if $\Omega_{\rm M}\simeq 0.3$ is the total matter fraction.

Actually, a more convenient framework is obtained by using comoving coordinates and the
superconformal time $s$ in place of the cosmic time $t$. We first rename $(\br,\bv)$ to
$(\br_{\rm phys},\bv_{\rm phys})$ and $\rho$ to $\rho_{\rm phys}$; then we set
\begin{equation*}
  \br_{\rm phys} = a \br \;,\quad \bv_{\rm phys} = 
  H\br_{\rm phys} + \frac{\bv}a \;,\quad H \equiv \frac{\dot a}a
  \;,\quad \rho_{\rm phys} = a^{-3}\rho \;,\quad dt =  a^2\, ds
\end{equation*}
where $a$ is the scale factor, solution of the acceleration equation of uniform
expansion
\begin{equation}\label{eq:a}
  \frac{\ddot a}{a} = \frac{4\pi G}3\left(\frac{\rhoM}{a^3} -2\,\rho_\Lambda\right)
  \;,\quad  \rhoM \equiv  \lowsub{\Omega}M \rho_{\rm crit} 
\end{equation}
and upper dots denote as usual derivative w.r.t. $t$.
In terms of the comoving $(\br,\bv)$ and the superconformal time $s$ 
the VP system of equations now reads
\begin{equation}\label{eq:VPcomov}
  \big[ \partial_s + \bv\cdot\nabla_r + (\nabla_r\phi)\cdot\nabla_v \big] f = 0\;,
\end{equation}
where $\phi$ is $a^2$ times the potential due solely to fluctuation over the
uniform background, that is
\begin{equation}\label{eq:phi}
  \begin{split}
    \Phi = \Phi_{\rm M} + \Phi_\Lambda &+ a^{-2}\phi \;,\quad
     \Phi_{\rm M} = \frac{2\pi G}{3a}\,\rhoM r^2 \;,\quad 
    \Phi_\Lambda = -\frac{4\pi G}{3}\,\rho_\Lambda a^2r^2\\
    \nabla_r^2\phi &= 4\pi G a \left[ \int d^3u\,f(\bx,\bu,s)- \rhoM \,\right]\;.
  \end{split}
\end{equation}
The time dependence of the scale factor if fixed by Eq.~(\ref{eq:a}), or by the equivalent
Friedmann equation in superconformal time
\begin{equation}\label{eq:aF}
  a^2 H =\frac1{a}\frac{da}{ds} = H_0\sqrt{\Omega_\Lambda a^4 +  \Omega_{\rm M}\, a}
\end{equation}
We see that, apart from the different symbol interpretation, the Liouville part of the VP
system is the same as before and the Universe expansion is equivalent to introducing a
negative matter background and time dependency on the gravitational coupling. The background
due to the cosmological constant drops out of the game and $\Omega_\Lambda$ directly
affects only the evolution of the scale factor.

\medskip 

To complete this dynamical setup, we have to provide the initial conditions for
eqs.~(\ref{eq:VPcomov}) and (\ref{eq:aF}). We assume that at $s=0$ the scale factor is $a
= a_{\rm i} = (1+z_{\rm i})^{-1}$, with the initial redshift $z_{\rm i}$ large enough so that DM is well
within the linear regime and the distribution function has to a very good approximation
the factorized form
\begin{equation}\label{eq:init}
  f(\br,\bv,s=0) = \rhoM \left[1+
    \delta_{\rm i}(\br) \right]\, f_{\rm i}(|\bv + \nabla \psi(\br)|)
\end{equation}
Here the overdensity fluctuation field $\delta_{\rm i}$ is a random realization of the
Gaussian process with the matter power spectrum at redshift $z_{\rm i}$, while $f_{\rm
  i}(|\bv|)$ is the freezed--out distribution, that is the unit--normalized space--uniform
isotropic velocity distribution at decoupling which would remain constant in the uniform
Universe, that is if $\delta_{\rm i}=0$. $\psi$ is the potential for the initial average
velocity field (the Zeldovich velocity) and satisfies
\begin{equation*}
  \nabla_r^2\psi =  a_{\rm i}^2\dot\delta_{\rm i} \simeq a_{\rm i}^2 H(a_{\rm i})\,
  \delta_{\rm i} = H_0\,\delta_{\rm i}
  (1+z_{\rm i})^{-1/2}\sqrt{\Omega_{\rm M} + \Omega_\Lambda(z_{\rm i}+1)^{-3}}
\end{equation*}
in order to fulfill to first order in $\delta_{\rm i}$ the mass continuity equation
\begin{equation*}
  \partial_s\rho + \nabla\cdot(\rho\avg{\bv}) = 0 \;,\quad
  \avg{\bv}(\br,s) \equiv \frac1{\rho(\br,s)}\int d^3v\,\bv\,f(\br,\bv,s)
\end{equation*}
at the initial time.

The specification of the matter power and of the freezed--out distribution $f_{\rm i}$
characterizes the type of DM. In the case of CDM the power spectrum has a slow power--like
falloff at small scales while $f_{\rm i}(\bv)$ is for all practical purposes a delta function at
$\bv=0$. In the case of WDM the power spectrum is more or less (depending on the specific
WDM model) sharply cut off at scales smaller than the DM free--streaming length, while
$f_{\rm i}(\bv)$ is a (highly model--dependent) isotropic distribution. Here we consider only a
specific model of WDM, namely fermionic thermal relics that decoupled at thermal
equilibrium while ultrarelativistic, so that, if $f_{\rm FD}$ denotes the dimensionless
Fermi--Dirac distribution, we have \cite{bode,hodal}
\begin{equation}\label{eq:f0FD}
   \rhoM \,f_{\rm i}(\bv) = q\,f_{\rm FD}(\epsilon) = q\,\frac{g}{1+e^{\epsilon/T}}
   \;,\quad q\equiv\frac{m^4}{(2\pi\hbar)^3}\;,
\end{equation}
where $\epsilon=mc|\bv|$ is the kinetic energy, $T$ is the (comoving) decoupling
temperature and $g$ is the number of non--translational degrees of freedom of the DM
particle, such as spin and/or internal symmetry quantum numbers. For example, $g=2$ or
$g=4$, depending on the specific model, in the case of a serious spin--1/2 candidate for
WDM such as the sterile neutrino \cite{stn1,stn2,stn3}.

From Eq.~(\ref{eq:f0FD}) we can derive the following expression for the freezed--out
velocity distribution ($v=|\bv|$ and $\zeta(x)$ is Riemann's $\zeta-$function):
\begin{equation}\label{eq:f0}
  \begin{split}    
    &f_{\rm i}(v) = \frac{A(B/\sigma_0)^3}{1 + e^{Bv/\sigma_0}} \;,\quad
    3\sigma_0^2 = 4\pi\int_0^\infty dv\,v^4f_{\rm i}(v) \\
    &A = \frac1{6\pi\,\zeta(3)} = 0.04413405\ldots \;,\quad 
    B = \left[\frac{5\,\zeta(5)}{\zeta(3)}\right]^{1/2} = 2.0768098\ldots\;,
  \end{split}
\end{equation}
with the identifications
\begin{equation*}
   \rhoM A \left(\frac{B}{\sigma_0}\right)^3 = g\,q \;,\quad T = \frac{m c\,\sigma_0}{B} \;.
\end{equation*}
Thus, recalling that $\rhoM=\Omega_{\rm M}\rho_{\rm crit}$, we obtain a rewriting
of the relation of refs.~\cite{bode,hodal} between the mass and the velocity dispersion of
the WDM particle (notice that $v_0=\sigma_0/B$ is most often quoted in the literature)
\begin{equation}\label{eq:sigma}
  \sigma_0 = 0.025 \left(\frac{h}{0.7}\right)^{2/3}
  \left(\frac{\Omega_{\rm M}}{0.3}\right)^{1/3}
  \left(\frac{2}{g}\right)^{1/3} \left(\frac{\rm keV}{mc^2}\right)^{4/3} {\rm km/s}\;.
\end{equation}
For what concerns the explicit choice of the WDM power spectrum, details are
provided in section \ref{sec:sphinit}.

\subsection{VP equation with spherical symmetry} \label{sec:sph}

The solution of the full six-dimensional VP equation presents a major challenge, both from
the analytical and numerical point of view. Analytical results are restricted to
perturbation theory and partial resummation thereof (see {\em e.g.} ref.~\cite{berna} and
references therein), while numerical results are only preliminary \cite{yoshi}.

The numerical study of the collapse of a single spherically symmetric system is much more
tractable, while retaining a great interest both theoretically and from the point of view
of its applications to concrete physical contexts. 

If the DM distribution function is rotational invariant, we can use as independent
phase--space variables $r=|\br|$, $u=(\br/r)\cdot\bv$ and the conserved
squared angular momentum (per unit mass) $\ell^2=|\br\wedge\bv|^2 = r^2\,(|\bv|^2-u^2)$. Then
\begin{equation}\label{eq:fnu}
  f = f(r,u,\ell^2,s)\quad,\qquad \rho(r,s) = \frac\pi{r^2}\int_{-\infty}^\infty
  du\int_0^\infty d\ell^2 \,f(r,u,\ell^2,s)
\end{equation}
and the VP equations read
\begin{equation}\label{eq:VPsph}
  \Big[ \partial_s + u\,\partial_r + 
  \Big(\frac{\ell^2}{r^3}-\phi'\Big)\,\partial_u \Big] f = 0
  \quad,\qquad \phi'(r,s) \equiv \partial_r \phi(r,s) = \frac{G\,a}{r^2}
  \left[M(r,s) - \frac{4\pi}3 \rhoM r^3\right] 
\end{equation}
where 
\begin{equation*}
  M(r,s) =  4\pi\int_0^r dr'\,r'^2\rho(r',s) \;, \quad
\end{equation*}
is the total dark mass within a sphere of radius $r$.  No derivative w.r.t. $\ell^2$ may
appear in the left hand of Eq.~(\ref{eq:VPsph}) because $\ell^2$ is a conserved quantity
in this collisionless dynamics.

\medskip

In this spherically symmetric framework, the moment expansion of the VP equation is
obtained by multiplying it by integer powers of $u$ and $\ell^2$ and then integrating over
$u$ and $\ell^2$. Using the standard notation of expectation values 
\begin{equation*}
  {\cal M}_{k,n}(r,s) = \avg{u^k\ell^{2n}} \equiv \frac{\int_{-\infty}^\infty
    du\int_0^\infty d\ell^2 \,u^k\ell^{2n}\,f(r,u,\ell^2,s)}
  {\int_{-\infty}^\infty du\int_0^\infty d\ell^2 \,f(r,u,\ell^2,s)} \,
\end{equation*}
and after one integration by parts in $u$, one obtains
\begin{equation}\label{eq:moments}
  \partial_t\left(\rho{\cal M}_{k,n}\right) + \frac1{r^2}\partial_r 
  \left(r^2\rho{\cal M}_{k+1,n}\right) = k\,\rho \left(\frac1{r}{\cal M}_{k-1,n+1}
    -\phi'{\cal M}_{k-1,n} \right)\;,\quad k,n=0,1,2\ldots \;,
\end{equation}
with the convention that ${\cal M}_{-1,n}\equiv 0$. When $k=n=0$ one has the mass continuity
equation with spherical symmetry
\begin{equation*}
  \partial_t\rho + \frac1{r^2}\partial_r\left(r^2 \bar{u}\rho\right) = 0 
  \;,\quad  \bar{u} = {\cal M}_{1,0} = \avg{u} \;,
\end{equation*}
while if $k=1$, $n=0$ and the continuity equation is used, one obtains Euler (or momentum)
equation
\begin{equation}\label{eq:euler}
  (\partial_t + {\bar u}\,\partial_r){\bar u} + \frac1\rho\,\partial_r(\rho\,\sigma_r^2)
  + \frac2{r}(\sigma_r^2 - \sigma_\theta^2) + \phi' = 0 \;,
\end{equation}
where
\begin{equation}\label{eq:sigdef}
  \sigma_r^2 = {\cal M}_{2,0}-{\cal M}_{1,0}^2 = \avg{u^2}-\bar u^2 \;,\quad 
  \sigma_\theta^2 = \frac{{\cal M}_{0,1}}{2 r^2} =  \frac{\avg{\ell^2}}{2 r^2}
\end{equation}
are the squared radial and tangential velocity dispersions, respectively. In the Cartesian
frame the pressure tensor reads
\begin{equation}\label{eq:pressure}
  P_{jk} = \rho\,\sigma_\theta^2 \delta_{jk} +  
  \rho \,(\sigma_r^2 - \sigma_\theta^2)\,\frac{r_jr_k}{r^2}
\end{equation}
and the vectorial form of Eq.~(\ref{eq:euler}) can be recovered by setting ${\bar u}_j =
{\bar u}r_j/r$.

In a quasi--stationary state with slowly varying hydrodynamic variables $\rho$ and $\bar
u$, Euler equation (\ref{eq:euler}) implies that wherever $\bar u=0$ we must also have
\begin{equation}\label{eq:jeans}
  \frac1\rho\,\partial_r(\rho\,\sigma_r^2)
  + \frac2{r}(\sigma_r^2 - \sigma_\theta^2) + \phi' = 0 \;,
\end{equation} 
which is known as Jeans equation in the astrophysical context. From the fluid point of
view, the condition $\bar u = 0$ is the marker of hydrostatic equilibrium.  On the other
hand, for the gas of individual particles orbiting in the slowly varying potential $\phi$,
the definition of {\em dynamic} equilibrium is perhaps more appropriate. Only in the
isotropic limit $\sigma_r^2 =\sigma_\theta^2\equiv P/\rho$, $P_{jk}=P\delta_{jk}$, Jeans
equation reduce to the equation of simple hydrostatic equilibrium $\partial_rP=-\rho
\phi'$.

\medskip 

To be useful, the infinite hierarchy in Eq.~(\ref{eq:moments}) needs to be approximately
closed at some order, but with no obvious physical basis for collisionless DM. On the
other hand, if effective collisional terms were added on some physical grounds, the
induced dissipation would allow to write an effective equation of state to relate the
pressure (then quickly rendered isotropic by dissipation) to the density and perhaps the
local entropy. DM is collinsionless and only more complex mechanisms, such as the effects
of localized DM clumps acting like macro--particles, could allow reliable closures at
higher order such as those in ref.~\cite{lapi}. Numerous clumps at many small scales are
indeed present in CDM bottom--up clustering, but they most likely do not play an
important role for WDM, due to the much smoother initial conditions. Tackling the full
hierarchy, that is the original VP equation, appears then mandatory for WDM.

\medskip 

The initial conditions in Eq.~(\ref{eq:init}) take now the form
\begin{equation}\label{eq:sphinit}
  f(r,u,\ell^2,s=0) =  \rhoM\, \left[1+
    \delta_{\rm i}(r) \right]\, f_{\rm i}\big(\sqrt{[u-u_{\rm i}(r)]^2+\ell^2/r^2}\,\big) \;,
\end{equation}
where $u_{\rm i}(r)$ is the initial infall velocity 
\begin{equation}\label{eq:u0}
  u_{\rm i}(r) = -\frac{\delta\! M_{\rm i}(r)}{4\pi\rhoM\, r^2}\,
  (1+z_{\rm i})^{-1/2}\sqrt{\Omega_{\rm M} + \Omega_\Lambda(z_{\rm i}+1)^{-3}}
\end{equation}
written in terms of the overmass
\begin{equation}\label{eq:oM}
  \delta\! M_{\rm i}(r) = 4\pi \rhoM \int_0^r dr'\,r'^2\,\delta_{\rm i}(r') \;.
\end{equation}
We see that the spherical symmetry has allowed to reduce the problem to three
dimensions. Of these, two dimensions correspond to the radial phase space with coordinates
$(r,u)$. The third coordinate $\ell^2$ enters the VP equation in (\ref{eq:VPsph}) as a
constant parameter, but is to be integrated over to obtain the density as in
Eq.~(\ref{eq:fnu}).

\section{Numerical setup}

Let us consider the Vlasov equation (\ref{eq:VPcomov}) regarding first $-\nabla\phi$ as an
external time--independent acceleration field. This is just the Liouville equation for the
incompressible streaming of any conserved local quantity in the phase--space of a single
particle. The formal solution can be written
\begin{equation*}
  f(s) = e^{s L} f(0) \;,\quad L = L_r + L_v \;,\quad 
  L_r = -\bv\cdot\nabla_r \;,\quad L_v = (\nabla_r\phi)\cdot\nabla_v  
\end{equation*}
where $L$ is the so--called Liouvillian operator. Of course $L_r$ and $L_v$ do
not commute and $e^{s L}$ cannot be factorized into the product of exponentials
of $L_r$ and $L_v$. But we can certainly write
\begin{equation}\label{eq:tau}
  e^{s L} = [\ e^{\tau L}]^n \;,\quad s = n\tau 
\end{equation} 
and for $\tau\to0$ 
\begin{equation}\label{eq:osplit}
   e^{\tau  L}  =   e^{\tau  \tilde L(\tau)} + O(\tau^3) \;,\quad 
  e^{\tau \tilde L(\tau)} \equiv e^{\tau L_v/2} e^{\tau L_r} e^{\tau L_v/2}
\end{equation}
The cubic order or the approximation can be checked by brute force power expansion of the
exponentials or, more simply, by noticing that
\begin{equation*}
    e^{-\tau  \tilde L(\tau)} = \big[e^{\tau  \tilde L(\tau)}\big]^{-1} =  
    e^{-\tau L_v/2} e^{-\tau L_r} e^{-\tau L_v/2} = e^{-\tau  \tilde L(-\tau)}
\end{equation*}
implies
\begin{equation*}
  \tilde L(\tau) = \tilde L(-\tau) = L +  O(\tau^2) \;.
\end{equation*}
Furthermore it should be noticed that 
\begin{equation*}
   e^{s  \tilde L(\tau)} = e^{-\tau L_v/2}\big[e^{\tau L_v} e^{\tau L_r}\big]^n  e^{\tau L_v/2}
\end{equation*}
and therefore one needs to alternatively apply many times $e^{\tau L_v}$ and $e^{\tau
  L_r}$, while $e^{\pm\tau L_v/2}$ only once at the beginning an the end. Notice also that
we may exchange the role of $L_r$ and $L_v$ in $\tilde L(\tau)$ without any problem. This
yields a second approximate evolution whose proximity to the first one can be used to
check the accuracy of both methods.

\medskip 

The promotion of $-\nabla\phi$ to a time--dependent acceleration field that depends on $f$
itself is rather straightforward: we need just to calculate $-\nabla\phi$ from the Poisson
equation just before every time $e^{\tau L_v}$ is applied. In the application to the
cosmological context of eqs.~(\ref{eq:VPcomov}) and~(\ref{eq:phi}) the gravity strength is
also growing in time with the scale factor $a$. We found that better stability in the
evolution is obtained by choosing non--uniform $s-$steps corresponding to uniform steps in
$a$, as can easily be determined from Eq.~(\ref{eq:aF}).

\medskip 

The great advantage of the well known operator splitting \cite{levecque} defined in
Eq.~(\ref{eq:osplit}) is that the half-step evolution operators, $e^{\tau L_v}$ or
$e^{\tau L_r}$, are just $\br-$dependent translations in $\bv$ or $\bv-$dependent
translations in $\br$. With the so--called finite volume methods \cite{levecque} these
translations (which correspond to the so--called {\sl advection} equation) can be
implemented very accurately on uniform as well as non--uniform grids on phase space, in
such a way to ensure machine--precision local conservation of the $f$ values at each
update. That is, the ${\cal L}_1$ norm of $f$ in any portion of the grid changes only
because of the flows at its boundaries. On the other hand, the ${\cal L}_n$ norms with
$n>1$, which are all exactly conserved in the continuum, are only approximately conserved
on any finite grid, unavoidably implying a numerical coarse graining. In the continuum
also the the squared modulus of the $f$ Fourier transforms in $\br$ and/or $\bv$, is
conserved. This ceases to apply after discretization, unless one makes use of the
non--local translation algorithm based on the discrete fast Fourier transform. This
algorithm, however, is more demanding in terms of computer power and is essentially
restricted to uniform grids; a serious limitation, as we shall see, in the present case of
gravitational collapse.

\medskip
Generally speaking all Vlasov solvers, that is numerical methods that solve directly the
Vlasov equation on a phase--space grid, are free of the noise inherent to $N-$body
simulations but suffer to one degree or another of numerical artifacts, like diffusion and
dissipation. In ref.~\cite{filbet} a comparison of different solvers was performed in the case
of a two--dimensional phase space. No method was found as a clear winner, especially when
the dependence of the acceleration on the density is considered, as necessary in the VP
system (in \cite{filbet} the context is that of plasma physics, but there a close
similarity to the cosmological setup in comoving coordinates). In fact, the
semi-Lagrangian methods with large integration time steps, which are certainly less
diffusive of those based on operator splitting plus finite volumes, are not allowed when
the acceleration field changes with time in a way that depends on the solution itself.

Altogether, provided diffusion and dissipation are under control, the simplicity and
stability of operator splitting and finite volume methods makes them particularly suited
for the purposes of this work, All numerical results reported in the sequel as based on
such a method. In particular, we have chosen a high--resolution finite volume advection
solver based on the piecewise linear reconstruction with total--variation--decreasing slope
limiters \cite{levecque}.  A similar choice was made in ref.~\cite{yoshi}, where the first
(low resolution) simulations in the full six--dimensional phase space are reported.

\subsection{Spherical symmetry and three--dimensional grids}

As already stated above, in this work we restrict our attention to the spherical
symmetric case of section \ref{sec:sph}. The phase space is then two--dimensional and,
together with the angular momentum $\ell^2$, we have a three--dimensional setup. Thus
high resolutions can be achieved. To obtain spherically symmetric initial conditions, we
perform averages over angles of the initial distribution $f(\br,\bv,s=0)$ of
Eq.~(\ref{eq:init}) around suitably chosen points as discussed in more detail below.

Some caution is necessary in the treatment of the squared angular momentum $l^2$, which of
course needs to be discretize too. The actual values used, and in particular their total
number, must be chosen wisely depending on the $(r,u)$ phase--space grid (see subsection
\ref{sec:angmom}). The latter cannot really be a static uniform grid, since the
gravitational collapse spans too many dynamical scales. In this case, where the collapse
center is fixed beforehand, it is simpler to opt for a static non--uniform grid which gets
finer and finer near $r=0$ and $v=0$.  Of course, in a higher--dimensional situation when the
collapse centers are not known apriori, the most sensible choice would be to use some
adaptive mesh refinement scheme.

High resolution requires indeed non--uniform grids. The radial density grows by
several order of magnitudes near the origin and the detailed evolving structure of this
peak is just the subject of this study.  At the same time radial velocities becomes very
large near the collapse center and the grid must allow for them since the only conceivable
boundary conditions in velocity space are those of free outflow and no inflow. A grid with
a too small velocity cutoff would lead to unphysical mass loss. A non--uniform grid may
allow for a large velocity cutoff while keeping within reason the total number of cells
in the $u$ direction.

We used non uniform phase--space grids which have $n_r \times n_u$ cells, with $n_r$ ranging
from 400 to 700 and $n_u$ from 600 to 1000, to select the best compromise between
resolution and speed. These cells have increasing width both for increasing $r$ and $|u|$,
with the exponential laws $\Delta x_{j+1}=(1+\epsilon_x)\Delta x_j$ for $x=r,u$ and
$0.015<\epsilon_r<0.028$, $0.02<\epsilon_u<0.03$, both variable parameters in search of the
best compromise between resolution and speed. The narrowest cells in $r$, which
are adjacent to $r=0$, have typical widths around 2 pc and the narrowest cells
in $u$, which are adjacent to $u=0$, have typical widths around 2 m/s.
The grid setup in the $\ell^2-$direction is described next.

\subsection{Angular momentum}\label{sec:angmom}

The conservation of angular momentum plays a crucial role in the spherically symmetric
collapse and must be handled very carefully. The angular momentum content of the system is
fixed once and for all by the initial distribution $f_{\rm i}(r,u,\ell^2)$ in
Eq.~(\ref{eq:sphinit}) and we need to numerically perform the integral over $\ell^2$ in
Eq.~(\ref{eq:VPsph}) with a level of accuracy which is consistent with the chosen
$(u,v)-$grid. Since $f_{\rm i}(r,u,\ell^2,s)$ dies exponentially fast for large $\ell^2$, the
integral can be cutoff at some finite value maintaining the desired accuracy. Much trickier
is the discretization near $\ell^2=0$.
 
Indeed, from Eq.~(\ref{eq:fnu}) and  Eq.~(\ref{eq:f0FD}), we see that
\begin{equation}\label{eq:cond}
  \int_{-\infty}^\infty du\int_0^\infty d\ell^2 \,f_{\rm i}(r,u,\ell^2) = \frac{r^2}\pi
\end{equation}
while $f_{\rm i}(r,u,\ell^2,s)$ dies exponentially fast in $1/r^2$ for any $\ell^2>0$. Still,
the condition (\ref{eq:cond}) should be fulfilled as closely as possible on the cell
centers of the $r-$grid when the integral over $u$ and $\ell^2$ are replaced by finite
sums. Moreover, one should worry also about the higher $u-$moment versions of
(\ref{eq:cond}). The higher $\ell^2-$moment versions are not as worrisome, since
the various $\ell^2-$ components are coupled only through their sum.
 
In our mass--conserving finite--volume setup, the integral over $u$ is naturally replaced
by the sum over $u-$cells properly weighted by the cell widths. For the sum $\ell^2$ we
have more freedom but, for obvious computational reasons, this sum should be as small as
possible. To reach a convenient compromise, we adopted a Gaussian quadrature scheme with
few free parameters which were optimized to fulfill Eq.~(\ref{eq:cond}) to a given order
for all cell centers of the $r-$grid. For grids with 400 $r-$cells with the first cell,
adjacent to $r=0$, of width $\sim 2$ pc, uniform agreement to order $10^{-3}$ requires a
costly $\ell^2-$grid with several thousands of points, where $\ell^2=O(1)$ pc$^2$
(m/s)$^2$ is the smallest value, that is few times smaller than the angular momentum
discretization scale corresponding to the $(u,v)-$grid.

Fortunately, the $\ell^2-$grid can be substantially coarsened by the following simple
strategy: as far as the initial conditions are concerned, we can relax the accurate
fulfillment of Eq.~(\ref{eq:cond}) and start with a coarser $\ell^2-$grid by including in
Eq.~(\ref{eq:init}) a suitable compensating factor $\kappa(r)$; $\kappa(r)$ cannot anyway
be too different from unity, since it allows to preserve the initial overmass, but cannot
prevent the inaccuracy in higher $u-$ and $\ell^2-$moments of $f_{\rm i}$. Of course this
trick is not possible during the evolution, since we do not know beforehand the r.h.s. in
Eq.~(\ref{eq:cond}) when $f(r,u,\ell^2,s)$ replaces $f_{\rm i}(r,u,\ell^2)$ in the l.h.s.;
we just have to run few prototype cases with finer and finer $\ell^2-$grids, with
correspondingly $\kappa(r)$ closer and closer to unity, until the evolution
stabilizes. Typically, this happens more or less when the number of $\ell^2$ values
reaches that of $r$ values. The corresponding $\kappa(r)$ exceeds unity by few percents
for all but the first and/or second leftmost $r-$grid center, where it can reach values up
to 10. In spite of this we could verify in the prototype cases, by measuring against long
runs with thousands of $\ell^2$ values, that the numerical evolution remains quite
accurate even on those cells.

\subsection{Boundary conditions}\label{sec:bc}

Boundary conditions must be set also at $r=0$ and at some finite value $R_{\rm max}$ of
$r$. On the line $r=0$ the correct condition is of reflecting type, with each half
$r-$line at a given $u>0$ and the corresponding half $r-$line at $u<0$ ``glued'' together
so that $r-$ translations act smoothly at $r=0$. Finally, at $r = R_{\rm max}$ we assume
free outflow, while the inflow can be null or free. The last two alternatives are actually
quite different, since with no inflow the total mass can only decrease with time, while
with free inflow it might go both ways depending on the initial conditions.  In the
cosmological context free inflow appears more natural, since the system under study is
just a very small portion of an initially nearly uniform Universe. 

On the other hand, owing to the presence of the background mass in Eq.~(\ref{eq:VPsph}),
it is possible to essentially decouple the DM halo from the rest of the Universe, thus
rendering the two alternatives of null or free inflow almost indistinguishable. To this
end, it is sufficient to set up the initial fluctuation profile $\delta_{\rm i}(r)$ in
Eq.~(\ref{eq:sphinit}) in such a way that the overmass $\delta\! M_{\rm i}(r)$ of
Eq.~(\ref{eq:oM}) vanishes at some point $r=R_0<r_{\rm max}$. Then the initial
infall velocity $u_{\rm i}(r)$ also vanishes at $r=R_0$ and in the first stages of evolution DM
flow away from $R_0$ in both directions. For the rest of the evolution till redshift zero,
a small but rather stable outward flux for $R_0<r<R_{\rm max}$ is maintained by the
gravitational push in the under dense region and by the free outflow conditions at $r=R_{\rm
  max}$. This prevents in a physical way any uncontrolled inflow at $r=R_{\rm max}$ when
inflow is allowed by the boundary conditions.

\subsection{Initial configurations}\label{sec:sphinit}

As anticipated at the end of section \ref{sec:VPDM}, we assume for the initial velocity
distribution $f_{\rm i}(v)$ the Fermi--Dirac form of Eq.~(\ref{eq:f0}) proper for fermions.
that decoupled at equilibrium while ultrarelativistic. $f_{\rm i}(v)$ parametrically
depends only on the initial one--dimensional velocity dispersion $\sigma_0$ given by
Eq.~(\ref{eq:sigma}). Typically, in our simulations we made the most straightforward
reference choice $h=0.7$, $\Omega_{\rm M}=0.3$, $g=2$ and $m=1\,$keV, so that
$\sigma_0=25\,$m/s.

\medskip

To complete the determination of $f(r,u,\ell^2,s=0)$ in Eq.~(\ref{eq:sphinit}), we need
instances of the initial overdensity profile $\delta_{\rm i}(r)$, from which also the initial
infall velocity $u_{\rm i}(r)$ can be determined according to Eq.~(\ref{eq:u0}). One possibility
is to make use of the reknown results of ref.~\cite{bbks} on the local peak statistics of random
Gaussian fields. We prefer here to proceed in a purely numerical fashion, to avoid the
averaging of local peak profiles over all random field realizations.
 
We set the initial redshift as $z_{\rm i}=100$ (and verified that $z_{\rm i}=200$ led to almost
indistinguishable results) and then followed the following procedure:
\begin{enumerate}
\item Compute the WDM power spectrom $P(k)$ for fermionic thermal relics of the given mass
  ($1$ keV/$c^2$) at redshift $z=z_{\rm i}$, using (a slight modification of) the 2011
  CAMB package \cite{camb}. Analytic expressions like those in refs.~\cite{bode,ddvs2}
  could be used with almost unnoticeable effects on the simulations.
\item Extract a random Gaussian field $g(\br)$ over a cubic lattice (we used two lattices, one
  with $512^3$ points and a lattice spacing of 45 kpc, the other with $768^3$ points and
  a spacing of  22 kpc), with zero mean and unit (in the sense of Kronecker's delta)
  variance.
\item Fast-Fourier transform $g(\br)$, multiply the result by $P(k)$ and
  inversely Fast-Fourier transform this product to obtain one realization of the fluctuation
  field $\delta_{\rm i}(\br)$ on the original lattice;
\item Repeat the previous step multiplying the Fourier transform of $g(\br)$ by
  $W(kR)\sqrt{P(k)}$, where $W(kR)$ is the Fourier transform of the sharp 
  window function $[3/(4\pi R^3)]\theta(R-r)$, namely
  \begin{equation*}
    W(y) = \frac3{y^3} \; (\sin y - y \; \cos y)
  \end{equation*}
  and $R$ is chosen with an eye to the characteristic mass $(4\pi/3)\rhoM\,R^3$
  of collapsed halos (but see below for more); then inversely Fast-Fourier transform to
  obtain a smoothed $\delta_{{\rm i},R}(\br)$ realization of the fluctuation field. Obviously
  $\delta_{\rm i}(\br)$ and $\delta_{{\rm i},R}(\br)$ differ significantly only if $R$ is significantly
  larger than the WDM free--streaming length, which for a 1 keV$/c^2$ thermal relic can be
  calculated to be 185 kpc \cite{dvs3}.
\item Select local maxima of $\delta_{{\rm i},R}(\br)$ which are {\em prominent}, that is, which are
  higher than all other peaks within a distance $R'$ to be properly selected (see below).
\item Perform a spherical average of the unsmoothed $\delta_{\rm i}(\br)$ around these prominent
  maxima and then interpolate over the chosen $(r,u)-$grid to obtain the radially
  symmetric peaked profiles $\delta_{\rm i}(r)$ to use in Eq.~(\ref{eq:sphinit}) and
  Eq.~(\ref{eq:u0}).
\end{enumerate}

Some comments are required on the procedure just outlined. First of all let us stress that
the aim is not really at an accurate halo statistics, for which more sophisticated
approaches are needed. Rather, we try and produce several initial spherically symmetric
peak profiles that are at least compatible with the true initial conditions, which are
certainly not spherically symmetric. In this respects, certain common interpretations
require some adjustment, especially because of the higher smootheness of the WDM
fluctuation fields $\delta_{\rm i}(\br)$ w.r.t. those of CDM.

For instance, in our approach the relation between the smoothing scale $R$ and the mass
$\bar M(R) = (4\pi/3)\rhoM\,R^3$ of collapsed halos is just marginal. Only after the
collapse, at redshift $z=0$, we can measure the actual mass of each DM halo and this
typically turns out to be larger than $\bar M(R)$. In fact, to trust the spherical
symmetry approximation, it is necessary that a collapse center be sufficiently isolated,
bringing in the other distance scale $R'$ characterizing the prominence as in point 5
above. By its definition, $R'$ must be chosen quite larger than $R$, but not necessarily
as proportional to $R$. The mass of a collapsed halo is in between $\bar M(R)$ and $\bar
M(R')$ and depends on specific features, as shown in Sec. \ref{sec:profiles}, of
the initial overdensity profile $\delta_{\rm i}(r)$. The chosen value of $R$, which must be in any
case larger than the free--streaming length, sets an approximate lower bound on the halo
mass and allows for collapse centers that are not maxima of the unsmoothed fluctuation
field $\delta_{\rm i}(\br)$. Notice indeed that the initial profiles $\delta_{\rm i}(r)$ are often not
monotonically decreasing nor maximal at $r=0$, although a common feature of all profiles
is their overall decrease toward zero (which is the average of the fluctuation field) at
larger distances, as dictated by the spherical average and the prominence requirement. The
latter also implies that quite often $\delta_{\rm i}(r)$ becomes negative sufficiently far
away. In turns, this often causes the corresponding overmass $\delta\! M_{\rm i}(r)$ to vanish at
some slightly larger distance.

\medskip

\begin{figure}[ht]
  \begin{center}
    \includegraphics[width=15.cm]{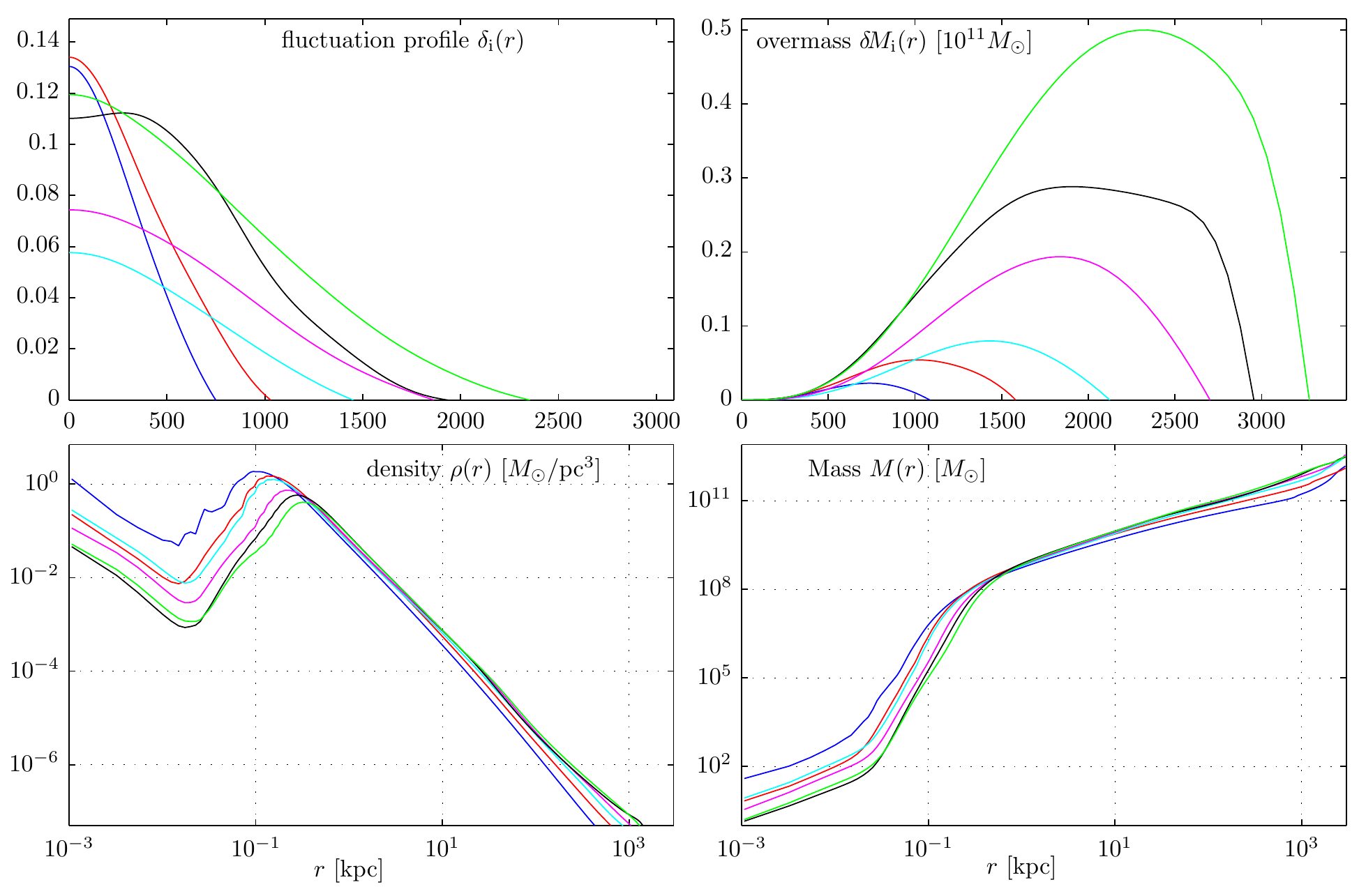}
    \caption{Upper panels: initial profiles, at redshift $z=100$, for six different
      realizations of the procedure described in Sec.~\ref{sec:sphinit}. Lower panels:
      density and mass at redshift $z=0$, after the collapse.}
    \label{fig:sixp}
  \end{center}
\end{figure}

In the upper panels of Fig.~\ref{fig:sixp} we show a small subsample of initial
fluctuation peaks obtained by the procedure outlined above, together with the
corresponding overmass. In the lower panel we anticipate the form of the collapsed
halos. The distance $R_0$ at which $\delta\! M_{\rm i}(r)$ vanishes determines
$M_0=(4\pi/3)\rhoM\,R_0^3$, the total amount of DM mass potentially involved in the
collapse, which in turns provides an upper limit to the mass of the collapsed halo for
whatever definition of such a mass one may choose. Sometime the overmass does not vanishes
within the simulation interval $0<r<R_{\rm max}$, in which case we modify by hand the far
end of the $\delta_{\rm i}(r)$, through the addition of a negative Gaussian peak centered
near $R_{\rm max}$, in order to determine a value of $R_0$ close to $R_{\rm max}$. As
explained in Sec.~\ref{sec:bc} the vanishing of $\delta\! M_{\rm i}(R_0)$ allows to
decouple the collapsing halo from the rest of the Universe, preventing on physical basis
a uncontrolled inflow rather than just setting the inflow to zero as a sharp (and
unphysical) boundary condition. On the other hand, all initial peak profiles for which the
arrangement by hand is necessary have essentially the same value of $M_0$. This feature,
together with selection of peaks by prominence (and probably several other reasons), quite
likely makes our sample of initial conditions statistically biased.

\subsection{On the CFL condition}

Named after Courant, Friedrichs, and Lewy, this is a necessary condition of stability (and
hence convergence) for any numerical method that aims at solving partial differential
equations (PDE). It can be formulated as ``{\sl A numerical method can be convergent only
  if its numerical domain of dependence contains the true domain of dependence of the
  PDE}'' \cite{levecque}. In the case of linear as well as non--linear transport
equations, this requirement is often translated into a bound of the form
\begin{equation}\label{eq:CFL}
  \nu \equiv \left|\frac{w\Delta t}{\Delta x}\right| \le  1
\end{equation}
on the so--called {\sl Courant number} $\nu$. Here $\Delta x$ represents the cell width of
a one--dimensional grid with a generic $x$ coordinate, $\Delta t$ is the integration
timestep and $w$ is the advection, or transport, velocity. To be more precise, there are in
general as many bounds like the one above as there are cells, and with time dependence
too, since $\Delta x$, $\Delta t$ and $w$ might all have a local nature, due to
non--uniformity of the grid, adaptive integration stepping and most notably dependence of
$w$ on the PDE unknowns when the transport equation is non--linear.

In our case $x$ is either $r$ or $u$, $w$ is one of the corresponding advection velocities
in Eq.~(\ref{eq:VPsph})
\begin{equation*}
  w_r = u \;,\quad w_u = \frac{\ell^2}{r^3}-\phi'(r,s) 
\end{equation*}
while $\Delta t$ is to be identified with the superconformal timestep $\tau$ of
Eq.~(\ref{eq:tau}). Then the bound in Eq.~(\ref{eq:CFL}) becomes prohibitively stringent
at small values of $r$ and/or large values of $u$, imposing values too small for $\tau$,
with uncontrollable diffusion and dissipation.  This would indeed constitute a major
obstacle to the whole approach if the bound on the Courant number was regarded as
necessary for the CFL condition to hold.

In reality, however, the bound of Eq.~(\ref{eq:CFL}) applies only to local updating
algorithms, that is methods in which the updated value of the variable of interest in a
given cell depends on the old values in the same cell and in a small, fixed
number (which depends on the specific method) of nearby cells (the so--called algorithm
stencil). The bound does not apply to non--local algorithms where the stencil grows with
the advection velocity.  

The simplest example is uniform advection in one dimension. In this case advection reduces
to rigid translation; then, if the lattice is uniform, by splitting the translation
parameter into an integer multiple of the lattice spacing $\Delta x$ plus the fractional
part, one can advect in a single step by amounts $w\Delta t$ much larger than $\Delta x$,
making the Courant number $\nu$ arbitrarily large. Using piecewise linear reconstruction,
it is not difficult to extend this method also to non--uniform lattices. Finally, since 
as explained above, the operator--splitting Vlasov solver reduces to intertwined
translations, we can forget about the bound of Eq.~(\ref{eq:CFL}) on the Courant number
and freely choose the superconformal timesteps in order to minimize diffusion and
dissipation while maintaining accuracy.

\subsection{Accuracy and stability tests}\label{sec:tests}

To check our algorithms against numerical artifacts we performed several stability and
accuracy tests on the core advection solver as well as on the full program. For brevity,
we report here only the main results of three full--program tests. Indeed, the advection
solver was adapted from quite standard, widely tested and employed algorithms (see
{\em e.g.}  ref.~\cite{levecque} and reference therein).

As accuracy quantifiers we considered the relative ${\cal L}_1-$norm, that is 
\begin{equation*}
  \Delta_X = \frac{\sum_kw_k\big|X_k^{(\rm m)}-X_k^{(\rm e)}\big|}{\sum_kw_k\big|X_k^{(\rm e)}\big|}
\end{equation*}
where $X$ is either the distribution $f$ itself or $M(r)$ (mass), $K(r)$ (kinetic energy),
$U(r)$ (potential energy) computed for the sphere of radius $r$. $X^{(\rm m)}$ stands for
the measured value while $X^{(\rm e)}$ stands for the one expected on theoretical
grounds. The sum runs over all indices of the observable, that is all grid indices in case
of $f$ or only the $r$ indices in case of $M$, $K$ and $U$. The weights $w_k$ are the
cell volumes or widths in the $r-$direction, respectively. The data reported in
Table~\ref{tab:tests} refer to a $n_r\times n_u \times n_{\ell^2}$ grid with $n_r=400$,
$n_u=600$ and $n_{\ell^2}=432$. Similar grids were used for most collapse simulations.

\bgroup
\def\arraystretch{1.5}%
\begin{table}
\begin{tabular}{c|c|c||c|c||c|c||c|c|}
  \cline{2-9}  
  & \multicolumn{2}{c||}{$ \Delta_M$}
  & \multicolumn{2}{c||}{$ \Delta_K$}
  & \multicolumn{2}{c||}{$ \Delta_U$}
  & \multicolumn{2}{c|}{$ \Delta_f$} \\
  \hline
  \multicolumn{1}{|c|}{test 1~~~~} &
  $ 2.8\!\times\! 10^{-4}  $&$ 8.5\!\times\! 10^{-3} $&$ 5.2\!\times\! 10^{-3} $&
  $ 1.7\!\times\! 10^{-2}  $&$ - $&$ - $&$ 2.3\!\times\! 10^{-2}  $&$ 0.1 $ \\
  \hline
  \multicolumn{1}{|c|}{test 2~~~~} &
  $\, 1.8\!\times\! 10^{-2} $&$\, 6.3\!\times\! 10^{-2} $&$\, 0.6 $&
  $\, 0.25 $&\, $5.7\!\times\!10^{-2} $&$\, 0.47 $&$\, 0.9\!\times\! 10^{-2} $&$\, 0.15 $\\
  \hline
  \multicolumn{1}{|c|}{test 3~~~~} &
  $\, 2.2\!\times\! 10^{-4} $&$\, 3\!\times\! 10^{-3} $&$\, 1.6 \!\times\!10^{-4} $&
  $\, 1.3 \!\times\! 10^{-3} $&$\, 2.8\!\times\! 10^{-4} $&$\, 1.2\!\times\!
  10^{-3} $&$ \, 7.5\!\times\! 10^{-2} $&$ 1.58 $ \\
  \hline 
\end{tabular}
\caption{Results of the test simulations described in Sec.~\ref{sec:tests}. For each
  quantifier the two column report the values at $t=T/10$ and $t=T$, except for test 3,
  for which they correspond to redshift $z=16$ and $z=0$.} 
\label{tab:tests}
\end{table}

\medskip
\begin{center}
  {\em Test 1: Free streaming}
\end{center}
This first test is just on free streaming, namely, without expansion nor gravity. The
trivial analytic solution $f(\br,\bv,t)=f(\br-\bv t,\bv,0)$ can be turned into a general
rule for the $f(r,u,\ell^2,t)$ of the spherical setup, but its implementation on the
$(r,u)-$grid requires rather intricate interpolations. We therefore consider an initial
distributions of factorized form $f(r,u,\ell^2,0)=\rho_{\rm i}(r) f_{\rm i}(v)$, so that
we immediately have the analytic form of $f^{(\rm e)}$, that is
\begin{equation*}
  f(r,u,\ell^2,t) = \rho_{\rm i}(r') f_{\rm i}(v)\;,\quad r'^2
  =  r^2 + 2ru\,t + \Big(u^2+\frac{\ell^2}{r^2}\Big)\,t^2 \;.
\end{equation*} 
In particular we take $\rho_{\rm i}(r)$ and $f_{\rm i}(v)$ to be both Gaussians with zero
mean and width $w_0$ and $\sigma_0$, respectively. As time span for the evolution we
take $T=\sqrt{3}w_0/\sigma_0$, which is the time needed for the spatial width to
double. From Table~\ref{tab:tests} one can see that the accuracy on $M$, $K$ and $U$
profiles is very good even if numerical diffusion is affecting $f$ quite substantially.
\medskip
\begin{center}
  {\em Test 2: King sphere}
\end{center}
This test checks the numerical stability of a King sphere, as in ref.~\cite{yoshi}, so
that the $X^{(\rm e)}$ are just the initial values of the $X^{(\rm m)}$.
In a notation slightly different from that in refs.~\cite{bt,yoshi}, we have
\begin{equation*}
  f(r,u,\ell^2,0) = C\,\theta(\psi)\,(e^\psi-1) \;, \quad \psi =
  \psi_0 - \frac1{2\sigma^2}\Big[u^2+\frac{\ell^2}{r^2} + 2\Phi(r)\Big] \;,
\end{equation*}
where $\theta(x)$ is the step function, $C$ is a constant that fixes the total finite mass
$M$ of the system, $\sigma$ is the central velocity dispersion, $\psi_0$ is King's shape
parameter and $\Phi(r)$ is the self--consistent potential. This ergodic phase--space
distribution is stationary and stable, but any numerical VP integrator inevitably changes
it. The smallness of the change, which necessarily grows in time due to numerical
diffusion and dissipation, is a measure of the accuracy and stability of the integrator.

To be definite, we set $\psi_0=8$ and choose the two scale parameters $C$ and $\sigma$ so
that, if $r_0$ is the King radius and $\rho_0$ is the central density, then
$r_0=4.77~\!$kpc and $\rho_0r_0=172.5\,M_\odot/$pc$^2$. The total mass is $M =
1.57\times10^{11}M_\odot$ and the tidal radius is $68.15\,r_0$. We take as time scale 
the Jeans free--fall time of the core, that is $t_{\rm J}=\sqrt{(\pi/(G\rho_0)}\simeq 0.01\,H_0^{-1}$.
and $T=10\,t_{\rm J}$ as time span for the simulation. 

The results in Table~\ref{tab:tests} show that coarse--grained observables suffers
differently from the large change of $f$ due to numerical diffusion. While the mass
changes only by few percents, the cumulative local change of the energies is ten
times as much. The behavior of $\Delta_K$, larger at $t=T/10$ than at $=T$, is due to
early fast diffusion in velocity space, which locally heats the system. These results show
that it is not wise to trust a numerical VP integration for times too large compared to the
natural time scale of the problem at hand. 

\medskip
\begin{center}
  {\em Test 3: Uniform expansion}
\end{center}

This third test consists in running a full cosmological simulation, from $z=100$
to $z=0$, of the uniform expanding Universe filled with WDM. That is, we set to zero the
initial fluctuations $\delta_{\rm i}=0$ and $u_{\rm i}=0$ in eqs.~(\ref{eq:sphinit}),
(\ref{eq:u0}) and (\ref{eq:oM}). This is manifestly a stationary solution of the
$6-$dimensional VP equation in comoving coordinates, Eq.~(\ref{eq:VPcomov}), since $\phi$
vanishes and $f(\br,\bv,s=0)$ in Eq.~(\ref{eq:init}) does not depend on $\br$. Of course,
also the spherically symmetric VP equation (\ref{eq:VPsph}) is satisfied, since the
isotropic freezed--out distribution $f_{\rm i}$ depends only on $v^2=u^2+\ell^2/r^2$ and
\begin{equation*}
  \Big[ u\,\partial_r + \frac{\ell^2}{r^3}\,\partial_u \Big] v^2 = 0
\end{equation*}
Hence the test essentially checks how accurately this equation numerically holds in our
program. Any error will be converted into small fluctuations, thus generating a fake
gravitational force which will in turn amplify the fluctuations, that is the numerical
errors. After all, it is at the heart itself of the theory of structure formation that
the uniformly expanding solution of the VP equation is unstable. Hence small perturbations
are amplified and at the linear level their characteristic amplitude grows proportionally
to the scale factor when matter dominates.

For this test $f^{(\rm e)}$ is just the initial $f$, $M(r)^{(\rm e)}$ is the background
mass $(4\pi/3) \rhoM r^3$ of Eq.~\ref{eq:VPsph}, while
\begin{equation*}
  U(r)^{(\rm e)} = -\frac{2\pi}{5}\rhoM a^2H^2r^5 \;,\quad  K(r)^{(\rm e)} =  -U(r)^{(\rm e)}
  +  2\pi \rhoM\frac{\sigma_0^2}{a^2} r^3 \;.
\end{equation*}
From the numbers in Table~\ref{tab:tests}, one sees that the physical instability of the
uniform expanding Universe causes a large change in $f$, as expected, but the effects of
fluctuations on mass and energy profiles do not grow beyond the linear level.

\section{Main results}\label{sec:results}

In this section we describe the most important features, according to our simulations, of
the gravitational collapse of the WDM halos. We selected many initial fluctuation profiles
according to the procedure in Sec.~\ref{sec:sphinit} and eventually collected 111
collapsed halos at redshift $z=0$. Our results show that basic quantities such as the
density and velocities or the virial ratio do stabilize to a large extent as $z\to0$ and
show universal properties. This quasi--stationary state is however quite different from
naive expectations based on approximate virialization or (local) thermalization arguments.

For simplicity we show only one--dimensional profiles, but the reader should keep in mind
that we computed the evolution in superconformal time $s$ of the full $3-$dimensional
distribution function $f(r,u,\ell^2,s)$. We also recall that the initial velocity
dispersion is $\sigma_0=0.025\,$km/s, consistent with a thermal relic mass
$m=1~$keV/$c^2$ according to Eq.~\ref{eq:sigma}. Reports on other values for $\sigma_0$
are provided only in Sec.~\ref{sec:sigma}.

\begin{figure}[ht]
  \begin{center}
    \includegraphics[width=15cm]{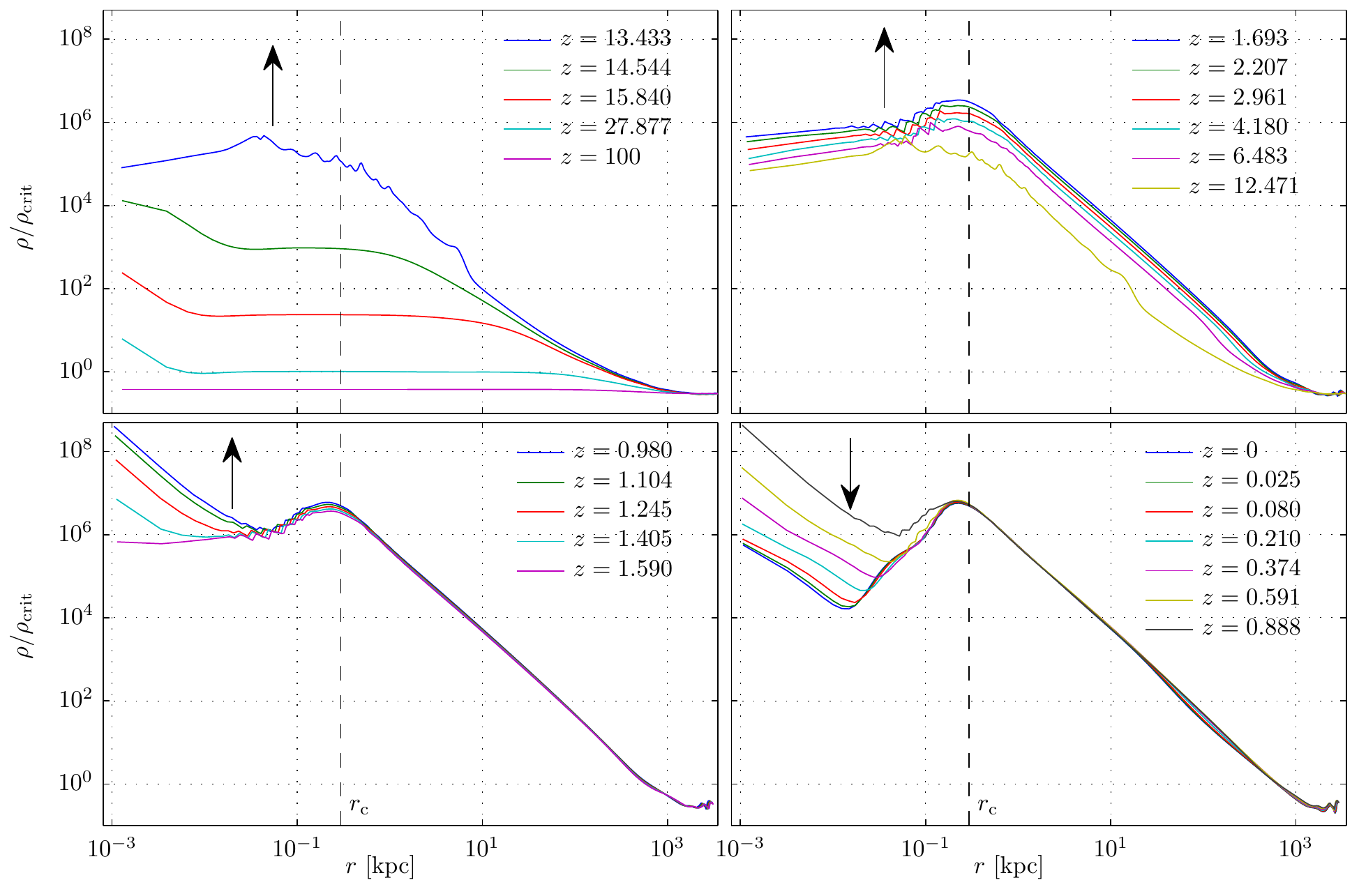}
    \caption{Evolution of the density profile in the halo example H1. The arrows emphasize
      the direction of change. The vertical dashed line indicate the position of the core
      radius $r_{\rm c}$ defined in Sec.~\ref{sec:rc}.}
    \label{fig:rho}
  \end{center}
\end{figure}

\subsection{Density and velocities}\label{sec:profiles}

An example of the evolution of the mass density profile $\rho(r)$ is depicted in
Fig.~\ref{fig:rho}. In this particular case the initial fluctuation profile $\delta_{\rm
  i}(r)$ was selected by prominence within $R'=3\,$Mpc among the highest ones in the
unsmoothed perturbation field. The peak maximum in the origin is $\delta_{\rm i}(0)=0.26$,
while $M_0=4.27\times 10^{12} M_\odot$. For ease of reference, let us name this example 
H1. We could regard H1 as a numerical approximation to the real, nearly spherical and
undisturbed collapse os a WDM halo.

Comparing the final density curves of H1 with those in the lower left panel of
Fig.~\ref{fig:sixp} one can appreciate the complete similarity, in spite of the large
difference in the initial peak heights and shapes. Indeed the cored and peculiarly hollow
shape of these plots is the crucial result of our simulations. This shape is universal, in
the sense that it is is common to all initial overdensity profiles considered. To
highlight this fact in the left panel of Fig.~\ref{fig:highres} we plot the density curve
of H1 superimposed to a set, denoted SH1, of 28 similar curves obtained from different
initial perturbations. Along the density curve the dots indicate the computation points,
that is the cell middlepoints.  In the right panel we plot the profiles, that is
$\rho/\rho_{\rm max}$ {\em vs.} $r/r_{\rm max}$, that show a good collapse onto a
universal shape.

\begin{figure}[ht]
  \begin{center}
    \includegraphics[width=15.cm]{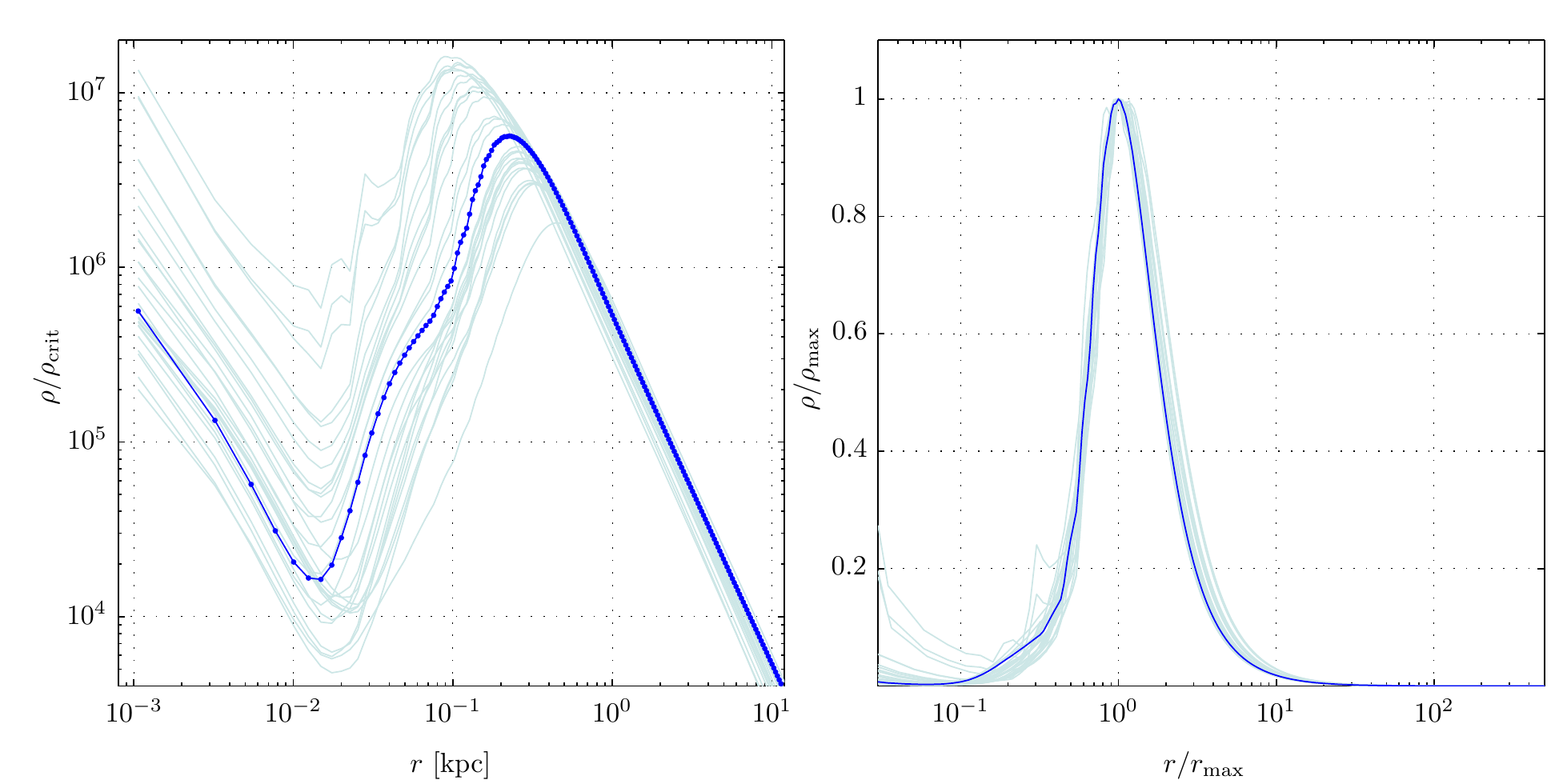}
    \caption{The $z=0$ density of H1 with the explicit indication of the computation
      points (left panel) and the scaled density (right panel). Paler lines correspond the
      the set SH1 with other 28 halos obtained from different initial conditions.}
    \label{fig:highres}
  \end{center}
\end{figure}

Indeed the presence of a maximum value $\rho_{\rm max}$ at a non--zero distance $r_{\rm
  max}$ is the most remarkable universal property of $\rho(r)$. The values of $\rho_{\rm max}$
and $r_{\rm max}$ vary instead from halo to halo, depending on the initial overdensity
profile that was used to generate them.

Another universal property is the nearly constant logarithmic slope in an extended region
on the right of the maximum. This slope stays very close to the value 2, that is $\rho(r)
\sim r^{-2}$, as can be verified by observing the presence of a clear plateau in the plot
of $r^2\rho(r)$ {\em vs.} $r$. The height of the plateau, as well as its
extent, are instead halo--dependent features. It is also quite evident that the plateau
extension, that is the region where $\rho(r) \sim r^{-2}$, is much wider than
in the halos of $N-$body simulations, a discrepancy most likely due our setup near the
boundary of the simulation box.

Clearly $r_{\rm max}$ already gives a natural definition of {\em core radius}, but we
will nevertheless provide a slightly different definition in the next section.

It must also be stressed that the history of $\rho(r)$ is very similar throughout our halo
sample, in spite of the differences in the initial $\delta_{\rm i}(r)$ profiles, with
essentially only two quantitative features that change from one halo to the other. The
first is the redshift of rapid formation of the maximum at $r_{\rm max}$ (a natural
definition of the moment of collapse), signaled by the steep rise in Fig.~\ref{fig:feat};
the second is the value $\rho_{\rm max}$ has at a fixed redshift before ($z=20$ in
Fig.~\ref{fig:feat}).  These two features are naturally strongly correlated.

The rather complex motion of the DM fluid can be appreciated in Fig.~\ref{fig:uavg}, where
the evolution of the average radial velocity $\bar u$ in H1 is depicted. Notice that $\bar
u$ vanishes as $z\to0$ in a whole region around the core radius $r_c\simeq r_{\rm max}$,
signaling the separation of the core from the outer more diffused halo. Namely, Jeans
equation (\ref{eq:jeans}) nearly holds around the core boundary and the core in is
equilibrium (hydrostatic in fluid viewpoint, dynamic in particle viewpoint) with the rest
of the halo.

On the other hand the development, near redshift $z=0$, of a large infall in the outer
part of the halo and an outward flux at even larger distances, show that the halo as a
whole is not in hydrostatic equilibrium. The outward flux leads to a slight loss of
matter.

\medskip

Also the radial and tangential velocity dispersions of Eq.~\ref{eq:sigdef} feature a quite
complex evolution, shown in Figs.~\ref{fig:sigr} and~~\ref{fig:sigt} again for halo H1.
One must notice the $r^{-1}$ decrease of $\sigma_\theta$ for $r>r_{\rm max}$ and its large
oscillation at small redshift in the central part of the halo. We find that the pressure,
Eq.~\ref{eq:pressure}), is strongly anisotropic, with the radial component $P_{rr}$ quite
smaller than the tangential component $P_{\theta\theta}$ in the region where the density
grows, a property that can be traced back to the conserved angular momentum. Both
components die faster than $1/r^2$ in the extended region outside the core where $\rho(r)$
dies as $1/r^2$ (and therefore the mass grows like $r$), since both velocity dispersions
also vanish as $r$ grows. This is not compatible with Jeans equation (\ref{eq:jeans}) but
is just in keeping with Euler equation (\ref{eq:euler}) with $\partial_t{\bar u}\approx 0$
and ${\bar u}\,\partial_r{\bar u} \not\approx 0$, due to the large infall in the outer
part of the halo.

\begin{figure}[ht]
  \begin{center}
    \includegraphics[width=10.cm]{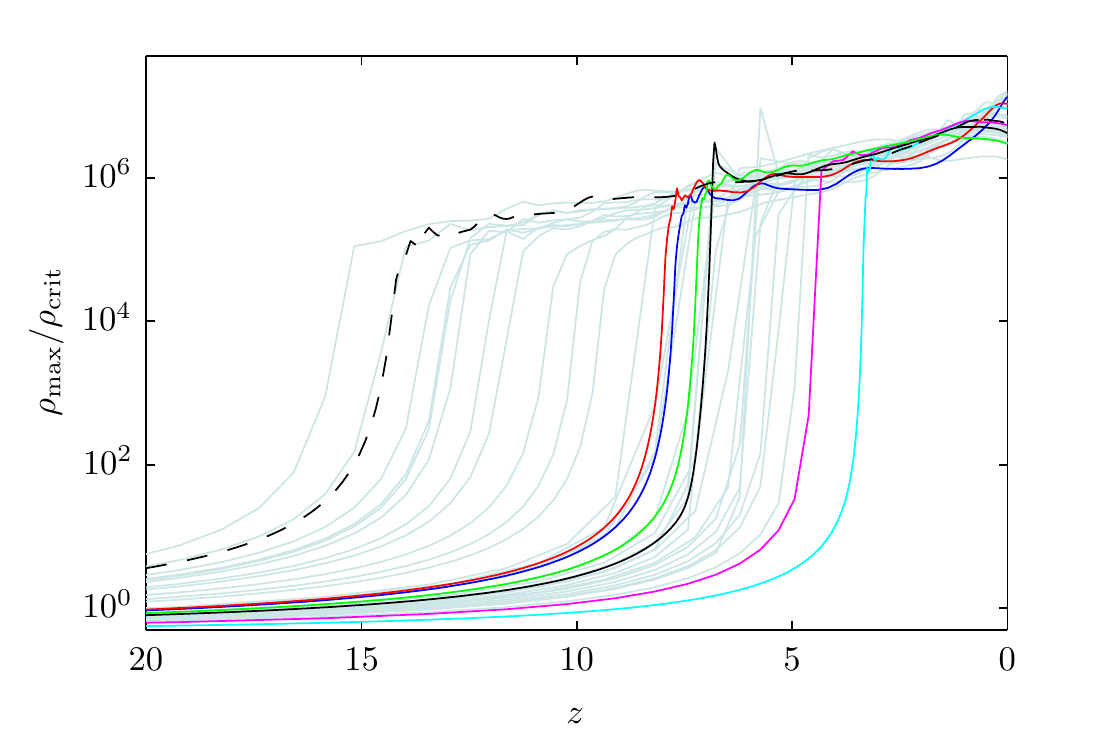}
    \caption{Evolution for $20>z>=0$ of the density value on the cell where
      $\rho=\rho_{\rm max}$ at $z=0$. The dashed line corresponds to H1. The lines in full
      color correspond to the 6 examples of Fig.~\ref{fig:sixp}. The paler lines
      correspond to the 28 halos of SH1; some of them are computed on a coarser grid of
      redshifts.}
    \label{fig:feat}
  \end{center}
\end{figure}

\begin{figure}[ht]
  \begin{center}
    \includegraphics[width=15cm]{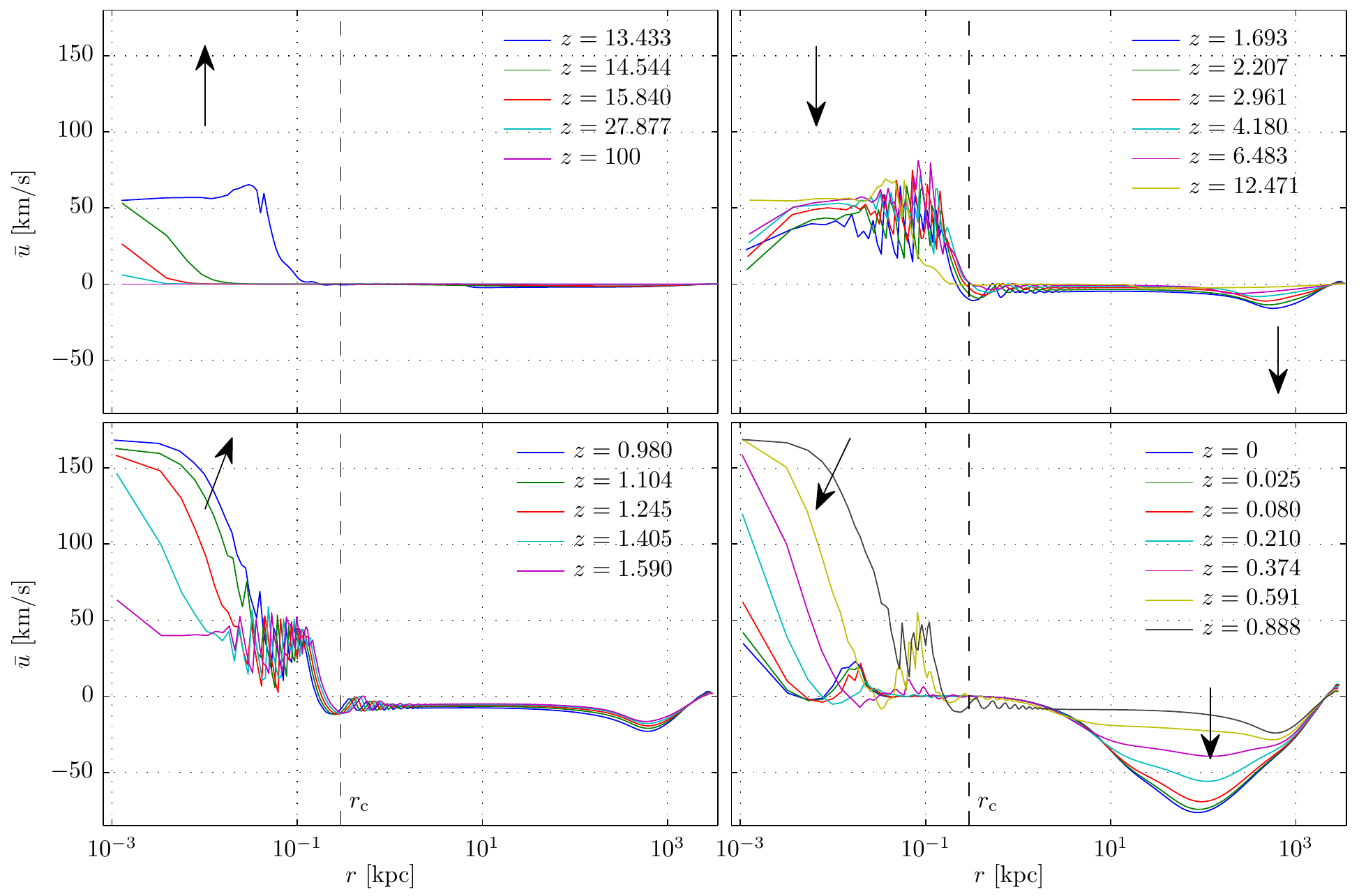}
    \caption{Evolution of the average radial velocity $\bar u\equiv \avg{u}$ in H1.  The
      arrows emphasize the direction of change. Notice the oscillations within the core
      and the large outer infall.}
    \label{fig:uavg}
  \end{center}
\end{figure}

\begin{figure}[ht]
  \begin{center}
    \includegraphics[width=15cm]{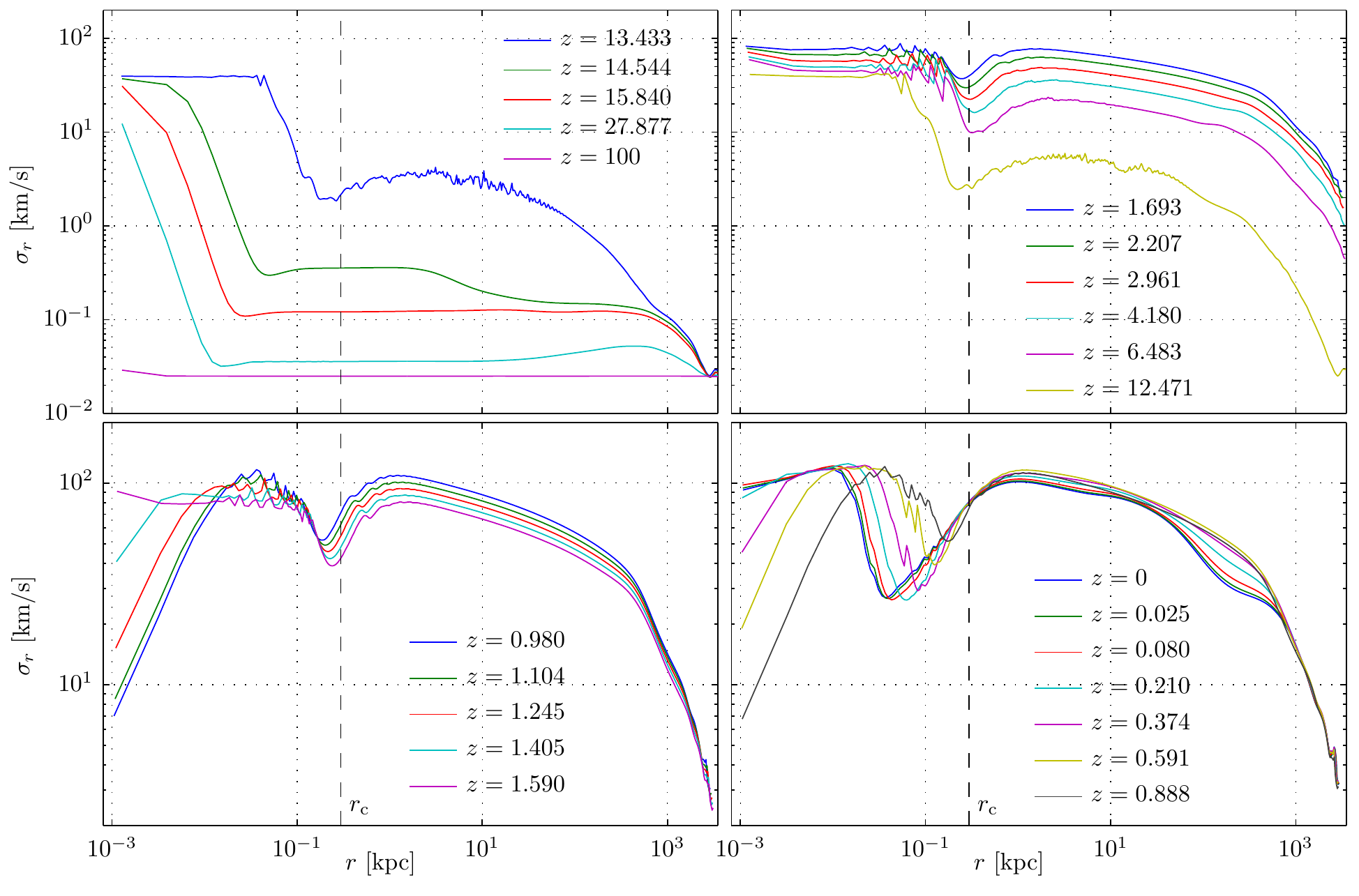}
    \caption{Evolution of the radial velocity dispersion $\sigma_r$ in H1. Notice the
      minimum inside the core and the rise across the region of maximal density.}
    \label{fig:sigr}
  \end{center}
\end{figure}

\begin{figure}[ht]
  \begin{center}
    \includegraphics[width=15cm]{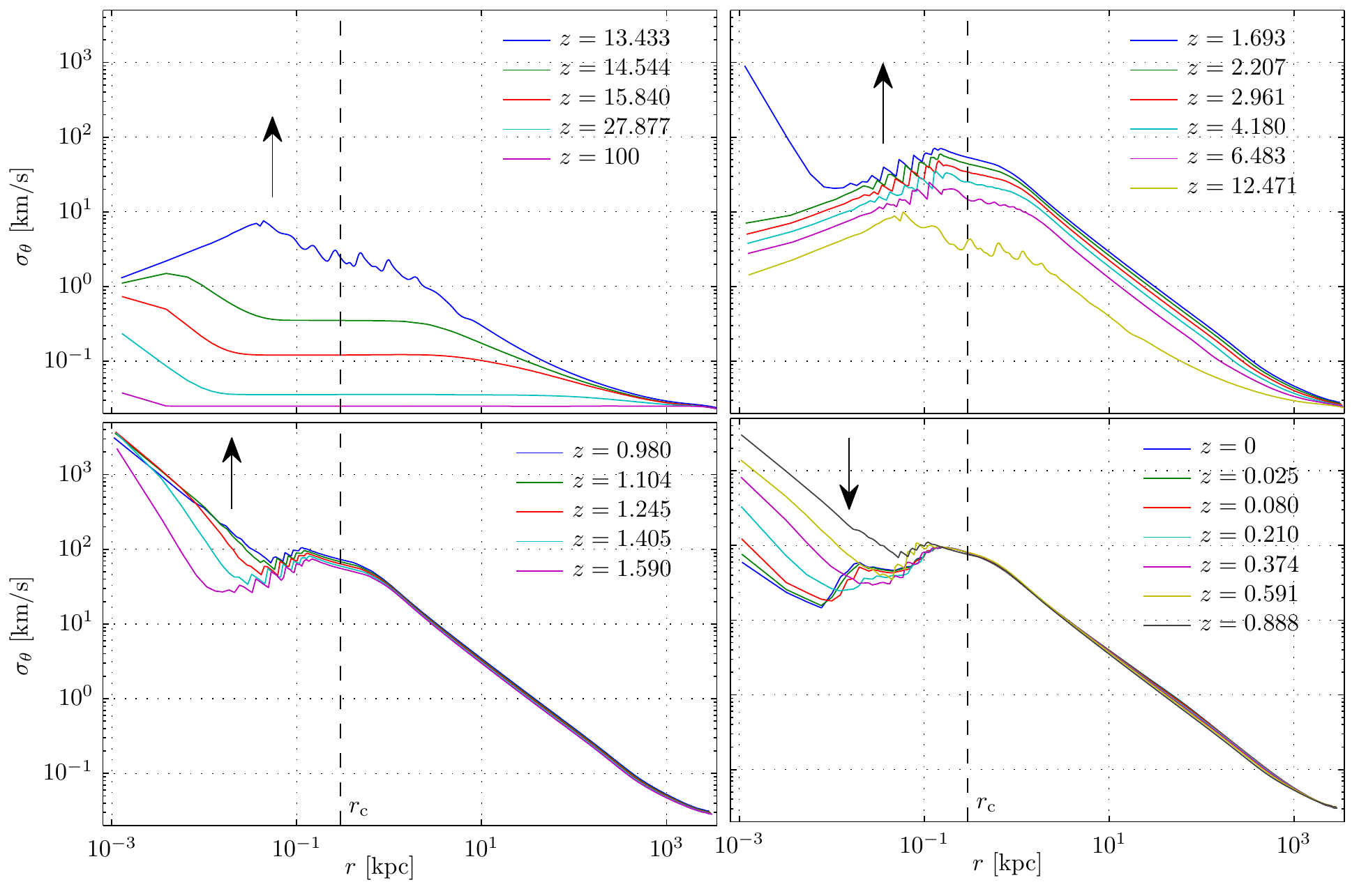}
    \caption{Evolution of the tangential velocity dispersion $\sigma_\theta$ in H1. The
      arrows emphasize the direction of change. Notice the correlation with the density
      evolution in Fig.~\ref{fig:rho}: when matter contracts it spins faster. }
    \label{fig:sigt}
  \end{center}
\end{figure}

\medskip

If gravitational effects remain the only means to observe DM, then the true DM observable of
our numerical halos is the mass $M(r)$ or, even better, the circular velocity $v_{\rm
  circ}$, related to the $M(r)$ by
\begin{equation*}
  v_{\rm circ}^2(r)  = \frac{G M(r)}{r} = \frac{4\pi G}3 r^2 \bar\rho(r) \;,
\end{equation*}
where $\bar\rho(r)$ is the mean density. Due to the hollow core, $\bar\rho(r)$ is not
everywhere larger than $\rho(r)$, as it would be if $\rho(r)$ were monotonically
decreasing as in simple equilibrium systems. In particular, $\bar\rho(r)$ is smaller than
$\rho(r)$ where $\rho(r)$ grows and larger than $\rho(r)$ beyond a certain point
$r_{\rm c}>r_{\rm rmax}$, where $\bar\rho(r_{\rm c})=\rho(r_{\rm
  c})\equiv\rho_{\rm c}$ [see the left panel of Fig.~\ref{fig:fits}]. These are universal
properties valid for all halos of our simulations.

By construction, $\rho_{\rm c}$ is the value at which the density should be cut to
replace the hollow core with a constant density core, that is to say,
\begin{equation*}
  M_{\rm c} = M(r_{\rm c}) = \frac{4\pi}3 \rho_{\rm c}r_{\rm c}^3
\end{equation*}
is exactly the mass of the core. But since $r\bar\rho'(r) = 3\left[\rho(r) -
  \bar\rho(r)\right]$, we see that $r_{\rm c}$ and $\rho_{\rm c}$ are also, perhaps more
simply, the position and the value of the maximum of $\bar\rho(r)$, respectively.

The circular velocity can now be rewritten as
\begin{equation}\label{eq:vc}
  v_{\rm circ}(r) = v_{\rm c} \frac{r}{r_{\rm c}}
  \left[\frac{\bar\rho(r)}{\rho_{\rm c}}\right]^{1/2} \;,\quad 
  v_{\rm c} \equiv v_{\rm circ}(r_{\rm c}) = \sqrt{GM_c/r_{\rm c}} \,.
\end{equation}
In the right panel of Fig.~\ref{fig:fits} we plot the profile $v_{\rm circ}(r)/v_{\rm
  c}$ {\em vs.}  $r/r_{\rm c}$ for H1 (blue curve) as well as the same 28 halos of
Fig.~\ref{fig:highres} (paler curves). The scaling across the core radius is very good and
a large region of nearly constant $v_{\rm circ}(r)$, a sort of plateau, can be identified,
although not really flat in the logarithmic scale. The plateau heights and shapes remain
halo--dependent also after the scaling by $v_{\rm c}$, but the height variation is very
limited in a neighborhood of 2. Most importantly, all plateaus are concave, and
particularly so where all profiles merge in the scaling region across $r/r_{\rm c}=1$ and
coalesce in the universal profile. On the other hand, for $r/r_{\rm c}<1$ this profile
drops faster than if the core were bulky rather than hollow, a consequence of the mass
deficit in the hollow core.

\begin{figure}[ht]
  \begin{center}
    \includegraphics[width=15cm]{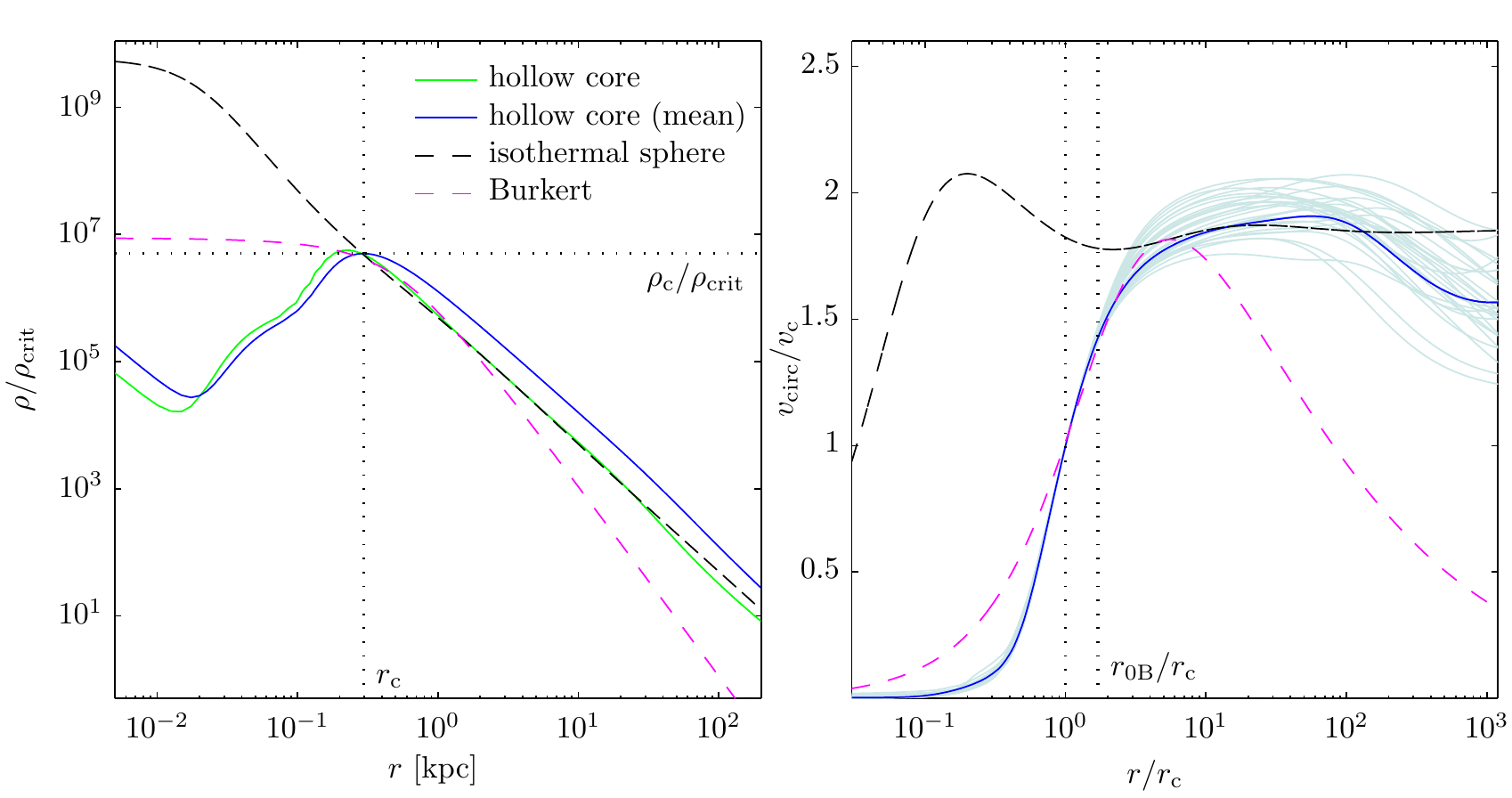}
    \caption{Left panel: comparison of the hollow density $\rho(r)$ of H1 with its mean
      $\bar\rho(r)$, with the density profile of the isothermal sphere and with the Burkert
      profile. Right panel: Circular velocity profiles of H1 (blue curve), of the same 28
      halos of SH1 (paler curves), of the isothermal sphere and of the Burkert
      profile. The isothermal sphere is a good fit for $r> 2\,r_{\rm c}$, the region
      explored by $N-$body simulations. The Burkert velocity profile is a good fit for
      $r_{\rm c}<r<10\,r_{\rm c}$. Notice that the Burkert core radius $r_{\rm 0B}$ is
      almost twice $r_{\rm c}$ as a direct consequence of the shape of the profiles.}
    \label{fig:fits}
  \end{center}
\end{figure}

\subsection{Core radius, surface density and other parameters}\label{sec:rc}

As anticipated by the adopted notation, we take $r_{\rm c}$ and $\rho_{\rm c}$ as
definitions of the {\em core radius} and {\em central, or core density}, respectively.

For H1, the halo under closer scrutiny, we find 
\begin{equation}\label{eq:rrhoc}
  r_{\rm c}=0.286\,{\rm kpc} \quad,\qquad \rho_{\rm c}=0.686\, M_\odot/{\rm pc}^3 
\end{equation}
which are values typical for the core of the DM halo in a dwarf galaxy.

\medskip

These values should be compared with the results of $N-$body simulations of
refs.~\cite{smco5,smco6}, which estimated in few pc the core radius of collapsed
halos produced by thermal relics with $m=1\,{\rm keV}/c^2$. Those simulations were run
with softening lengths of few hundreds pc and different values of initial velocity
dispersion $\sigma_0$, while keeping the initial conditions typical of
$m=2~$keV/$c^2$. [To maintain numbers under control, one has to recall that our $\sigma_0$
is the $1-$dimensional velocity dispersion, larger than the characteristic velocity
$v_0(z=0)$ quoted in ref.~\cite{smco5} by the factor $B=2.0768098\ldots$ of
Eq.~(\ref{eq:f0}).]  At values of $\sigma_0$ smaller than $0.025\,$km/s, the value
consistent with $m=1~$keV/$c^2$ and our reference choice, the cores were not resolved, but
the density profile was found very close to that of CDM for $r>1\,$kpc, that is for
$r>3\,r_{\rm c}$. The first resolved core was found for $m=0.13~$keV/$c^2$ (simulation WDM3
in ref.~\cite{smco5}) with a quoted core radius $r_{\rm core}=0.42\,$kpc. Using the data
from simulations at higher $\sigma_0$ and their proximity from below to the theoretical
$Q$-based upper bound \cite{hodal}, $r_{\rm core}$ was extrapolated to be $\lesssim
10\,$pc for $m\gtrsim 1~$keV/$c^2$.

\medskip 

The discrepancy with our result for H1, $r_{\rm core}=r_{\rm c}=356\,$pc is due the
extrapolation procedure (mis--)guided by the $Q-$based bound.  Indeed, the $Q-$based bound is
itself based on the apriori assumption mentioned in the Introduction, namely that the
isothermal sphere provides a good description of the halo core.

Suppose to try and fit the circular velocity $v_{\rm circ}(r)$ of H1 with that of an
isothermal sphere, in the hypothesis of knowing $v_{\rm circ}(r)$ only outside the core, say
$r>2\,r_{\rm c}$. The circular velocity $v_{\rm iso}(r)$ of the isothermal sphere with
King's radius $r_0$ and central density $\rho_0$ can be written as
\begin{equation*}
  v_{\rm iso}(r) = \sqrt{2}\,\bar\sigma g(r/r_0)
\end{equation*}
where $\bar\sigma$ is the isothermal one--dimensional velocity dispersion 
\begin{equation*}
  \bar\sigma = \frac13\sqrt{4\pi G \rho_0r_0^2} \;.
\end{equation*}
and $g(x)$ is the well known profile that performs dampened oscillations around unity
as $x\to\infty$. In the right panel of Fig.~\ref{fig:fits} we plot with a dashed line 
the profile $1.3\,g(x/15)$ {\em vs.} $x=r/r_{\rm c}$. 

It is natural to start by matching the H1 plateau value of $v_{\rm circ}(r)$, which has
roughly the value $1.8\,v_{\rm c}$, by setting $\sqrt{2}\bar\sigma \simeq 1.8\,v_{\rm c}$, or
\begin{equation*}
  \rho_0r_0^2 \simeq  4.86\,\rho_{\rm c}r_{\rm c}^2 \;.
\end{equation*}
It remains to fix the relative distance scale, that is $r_{\rm c}/r_0$. If the plateau
were really flat, the best fit would require the limit $r_0\to0$ yielding the singular
isothermal sphere, in order to flatten out the characteristic oscillations of $v_{\rm
  iso}(r)$ for large but finite $r/r_0$. Since the plateau is not flat, a better fit could
be obtained at some $r_0>0$, but certainly $r_0$ must be small enough so that at least the
first large oscillation of $g(r/r_0)$ gets out of the way, as in the right panel of
Fig.~\ref{fig:fits}. Since we assumed that $r=2\,r_{\rm c}$ is the closest we can get to
the core and $x\simeq30$ is the location of the first minimum of $g(x)$ on the right of
its maximum, we have
\begin{equation*}
  r_0 \lesssim r_{\rm c}/15 \;.
\end{equation*}
$15\,r_0$ is indeed the value at which the density $\rho_{\rm iso}(r)$ of the isothermal
sphere is declared to enter its mean $1/r^2$ regime \cite{bt}. In the left panel of
Fig.~\ref{fig:fits} we plot $\rho_{\rm iso}(r)$ with $r_0 = r_{\rm c}/15$ and $\rho_0 =
4.86\times 15^2 \rho_{\rm c}$.

Given reasonable mock data for the specific halo H1, the fit could probably be improved,
but such an exercise is beyond the scope of this discussion, which is to show that the
dashed black lines in Fig.~\ref{fig:fits} provide the natural extrapolations of density and
circular velocity profiles when the core is not resolved {\em and an isothermal sphere is
  assumed}. Using the values (\ref{eq:rrhoc}) of H1 we find
\begin{equation*}
  \rho_0 \gtrsim 750\,M_\odot/{\rm pc}^3 \quad,\qquad r_0\lesssim 20\,{\rm pc} \,,
\end{equation*}
in keeping with the extrapolated results of $N-$body simulations in
refs.~\cite{smco5,smco6}, whose quick conclusion was that keV--ranged WDM does not help in
solving the core problem of DM halos \cite{sch}. 

We rather conclude that, even when $N-$body simulations resolve the core by increasing the
initial velocity dispersion, they underestimate the final velocity dispersion in the core
region, since not so many ``particles'' with $10^5M_\odot$ mass can fit into a
$10^7M_\odot$ core. This implies an overestimation of the phase--space density $Q$ to values
close to the theoretical bound, with the proximity holding through the
misguided extrapolation. In Sec.~\ref{sec:sigma} we provide data on the relaxation factor
$Z$ of Eq.~(\ref{eq:Z}) which show that, on the contrary, also in the core region $Q$
starts dropping very rapidly as the initial velocity dispersion $\sigma_0$ is decreased
below $0.40\,$km/s, that is when $m>0.125\,{\rm keV}/c^2$.  

\medskip

With $r_{\rm c}$ and $\rho_{\rm c}$ we can compute the so--called {\sl central surface
  density} $\mu_{\rm c}=\rho_{\rm c}r_{\rm c}$. For the case at hand, example H1 with the
values as in Eq.~\ref{eq:rrhoc}, we find
\begin{equation}\label{eq:muc}
  \mu_{\rm c} = 203\, M_\odot/{\rm pc}^2 
\end{equation}
in remarkable, but rather puzzling agreement with the observed value $\mu_{\rm
  0,obs}=140^{+83}_{-52} M_\odot$/pc$^2$ \cite{surd2,surd4}. Puzzling because the observed
value is obtained by fitting the density of DM halos with the Burkert profile \cite{burk}
\begin{equation*}
  \rhoB(r) = \rhozB B(r/r_{\rm 0B}) \;,\quad 
  B(x) = \frac1{(1+x)(1+x^2)}
\end{equation*}
which is quite different from the hollow one. 

The surface density in refs.~\cite{surd2,surd4} is defined as $\mu_0=\rho_{\rm
  0B}r_{\rm 0B}$. It can be estimated for our hollow cores by trying and fit the circular
velocity Eq.~(\ref{eq:vc}) with $v_{\rm B,circ}(r)$, the circular velocity obtained from
the Burkert density, that is
\begin{equation*}
  v_{\rm B,circ}^2(r) = 2\pi G \mu_0 r_{\rm 0B}\, x^{-1}\!\left[\log(1+x) + \tfrac12
    \log(1+x^2) - \tan^{-1}(x) \right] \;, \quad x = r/r_{\rm 0B} \;.
\end{equation*}
The result is depicted by the magenta green lines in Fig.~\ref{fig:fits}. Notice that a
good fit is possible only in some finite interval on the right of $r_{\rm c}$, because of
the mass deficit of the hollow core and the too long $r^{-2}$ tail of the diffuse part of
the halo. In Fig.~\ref{fig:fits} the interval, fixed a priori, is $r_{\rm c} < r
<10\,r_{\rm c}$. A closer fit in a narrower interval is possible at the price of
increasing the gap between $V_{\rm B,max}$, the maximal Burkert circular velocity, and
$V_{\rm max}$, the actual maximum of $v_{\rm circ}$. Our choice is just a reasonable
compromise. The best fit values are
\begin{equation*}
   r_{\rm 0B} = 1.7\,r_{\rm c} = 0.504\,{\rm kpc} \;,\quad 
   \mu_0 = 2.97\, \mu_{\rm c} = 603\,M_\odot/{\rm pc}^2  \;,\quad 
   V_{\rm B,max} = 58.2\,{\rm km/s} \;,
\end{equation*}
to be compared with Eq.~(\ref{eq:rrhoc}), (\ref{eq:muc}) and $V_{\rm max}=63\,{\rm
  km/s}$. The concave shape and good scaling properties of $v_{\rm circ}^2(r)$, quite
evident from right panel of Fig.~\ref{fig:fits}, imply that $\mu_0/\mu_{\rm c}$ have
similar values in all other halos of our sample for any other reasonable choice of fit
interval. It is important to notice that $r_{\rm 0B}$ is almost twice $r_{\rm c}$, that is 
the fit is good also inside the Burkert core, down to almost $r_{\rm 0B}/2$. This internal
fitting further improves as the initial velocity dispersion $\sigma_0$ increases (see
Sec.~\ref{sec:sigma}).

$V_{\rm B,max}$ can be used to quantify the mass content of our halos, without any
reference to the excessively long $r^{-2}$ tail of the diffuse part. Thus the deviation of
$\mu_0$ from the observed value is a direct measure of their core concentration w.r.t.
real DM halos. In the core region, H1 is roughly four times more concentrated than
observed DM halos because it has $\mu_0/\mu_{\rm 0,obs}\simeq 4$, that is a maximal
circular velocity four times larger than the typical real DM halo with the same core size.
Moreover, since $\mu_{\rm c}$ happens to be very close to $\mu_{\rm 0,obs}$ (for no
obvious reasons, at the moment), we see that also the ratio $\mu_0/\mu_{\rm c}$ provides a
simple quantitative measure of the basic difference between the hollow--core H1 and real
DM halos.

This analysis is the least restrictive, of course, since we are fitting H1 with a Burkert
profile only where there the fit can be good, namely for $0.5\,r_{\rm 0B} \lesssim r
\lesssim 5\,r_{\rm 0B}$.  Since the Burkert profile allows very good fits to the circular
velocities of real DM halos down to few percents of $r_{\rm 0B}$, one could say that the
hollow--core halos of $1\,$keV WDM, at least as obtained in our simulations, are ruled out
by observations. Still, the closeness of $\mu_{\rm c}$ with $\mu_{\rm 0,obs}$ and the relative
smallness of $\mu_0/\mu_{\rm c}$ provide a very interesting starting point for further
improvements and enhancements.

\medskip

\begin{figure}[ht]
  \begin{center}
    \includegraphics[width=15cm]{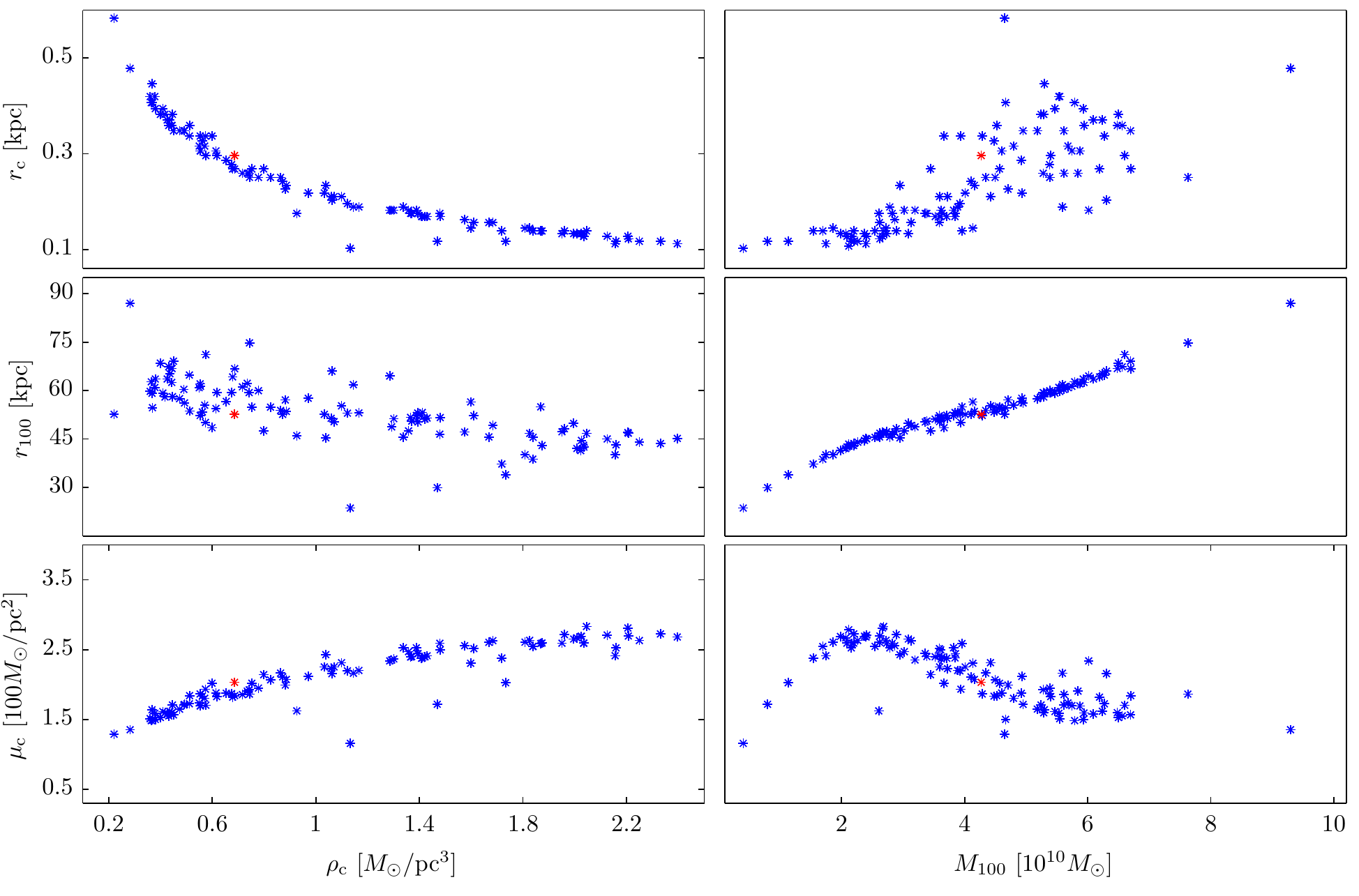}
    \caption{Scatter plot of the core parameters defined in Sec.~\ref{sec:rc} as computed
      in $111$ collapse simulations starting from different initial overdensity profiles
      obtained as in Sec.~\ref{sec:sphinit}. The red marks correspond to halo H1.}  
    \label{fig:splots}
  \end{center}
\end{figure}

To characterize also the diffuse part of our hales, we consider the pair $r_{100}$ and
$M_{\rm halo}=M_{100}$. These are the radius at which the density drops to the value $100\,
\rho_{\rm crit}$ and the mass of the halo contained within the corresponding sphere. Since
$r_{100}$ falls always within the region where $\rho(r) \sim r^{-2}$ and $M(r)\sim
r$, we expect $r_{100}$ and $M_{100}$ to be roughly proportional. We include $r_{100}$ and
$M_{100}$ for better completeness only, because we have reasons to believe that the
excessive length of the $r^{-2}$ tail is a spurious effect of our initial and boundary
conditions.

In Fig.~\ref{fig:splots} we show some scatter plots of the above parameters over the 111
halos of our sample. In particular, from the scatter plot of $r_{\rm c}$ vs. $\rho_{\rm
  c}$ in the upper left panel, one can appreciate more clearly the property already quite
evident from the lower left panel of Fig.~\ref{fig:sixp}, namely that profiles with a
smaller $\rho_{\rm c}$ have a larger $r_{\rm c}$.

The lower panels of Fig.~\ref{fig:splots} exhibits the scatter plot of $\mu_{\rm c}$
vs. $\rho_{\rm c}$ and vs. $M_{100}$.  We see that $\mu_{\rm c}$ grows quite slowly, by a factor
2, as the central density grows by more than a decade, while it decreases as $M_{100}$
varies over a decade.

The mean, median, minimal and maximal values of the surface density are, respectively
\begin{equation*}
  \mu_{\rm c}\frac{{\rm pc}^2}{M_\odot} \;: \; 213,216,116,283 \quad ,\qquad
  \mu_0\frac{{\rm pc}^2}{M_\odot} \;: \;   650,650,377,901 \;.
\end{equation*}
to be compared with the observed value $\mu_{\rm 0,obs}=140^{+83}_{-52} M_\odot$/pc$^2$
\cite{surd2,surd4}.  The ratio $\mu_0/\mu_{\rm c}$ oscillate between $2.89$ and $3.36$.
The slow decrease of $\mu_0$ with the halo mass $M_{100}$ disagrees with the slow increase
observed in recent data \cite{boya}, but our determination of $M_{100}$ most likely
suffers from the biased selection procedure of initial profiles.

\subsection{Potential and phase--space density}\label{sec:phiQ}

Fig.~\ref{fig:phi} shows the evolution of the ``reduced'' gravitational potential $\phi$
of Eq.~(\ref{eq:phi}). At redshift $z=0$, wherever the complete background density
$\rhoM-2\rho_\Lambda$ can be neglected w.r.t. to actual DM density, this potential
coincides with the full gravitational potential $\Phi$. From Fig.~\ref{fig:phi} one can
see that, in keeping with its coarser--grained nature and unlike the density and
especially the velocity profiles, the $\phi$ profile varies monotonically, smoothly
developing a well with a nearly flat bottom in $\log r$. This is another clear manifestation of
the hollow nature of the core.

\begin{figure}[ht]
  \begin{center}
    \includegraphics[width=10cm]{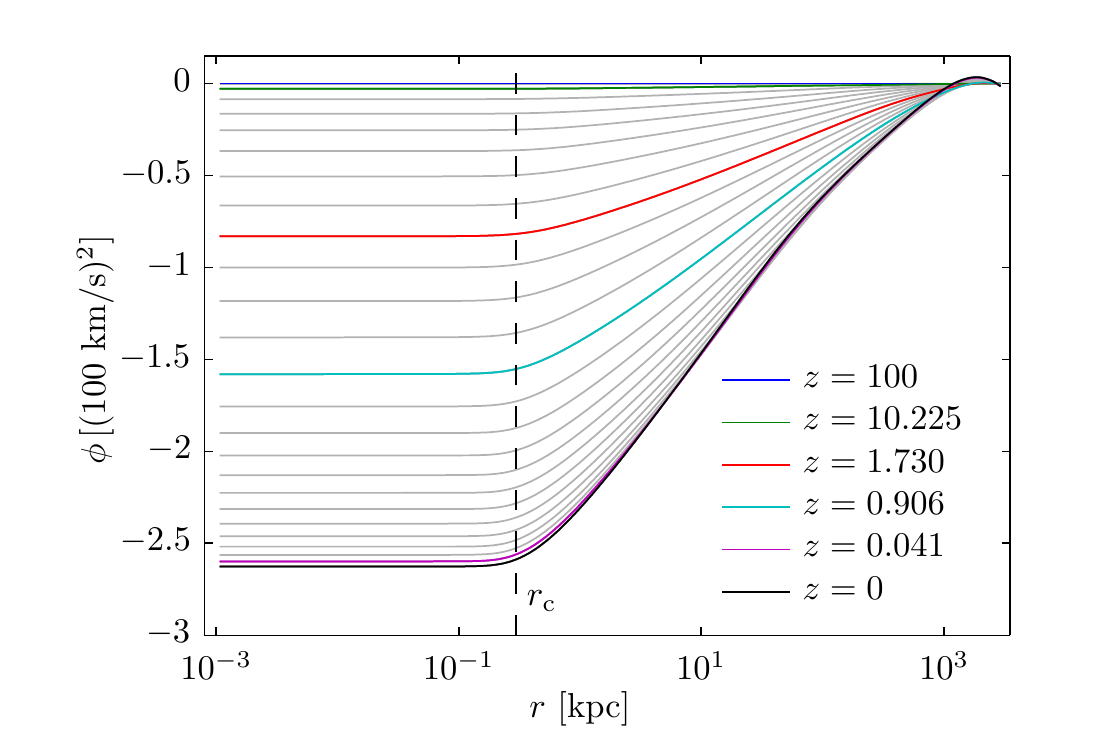}
    \caption{Evolution of the ``reduced'' gravitational potential $\phi$ in the case
      $\delta_{\rm i}(0)=0.26$. Gray lines correspond to intermediate redshifts.}
    \label{fig:phi}
  \end{center}
\end{figure}

The (pseudo) phase--space density is given by $Q=\rho/\sigma^3$, where $3\sigma^2 =
\sigma_r^2 + 2 \sigma_\theta^2$ is the total velocity dispersion. It provides an estimate
of the full distribution function for small velocities. In the purely classical context of
this work, where values of the distribution function $f$ are just transported through
phase space by the VP equation, any coarse--grained approximation of $f$ cannot increase
during the collapse. Rather, the larger its decrease, the stronger the violent relaxation
associated with the collapse (see Sec.~\ref{sec:sigma}). Even if $Q$ is not exactly a
coarse graining of $f$, but a ratio of coarse--grained quantities that scales like $f$, it
is expected to behave similarly.

In Fig.~\ref{fig:Z} we plot an example of the evolution of (the local version of) the
relaxation factor \cite{dvs}
\begin{equation}\label{eq:Z}
  Z(r,z) = \frac{Q(r,z_{\rm i})}{Q(r,z)} \;.
\end{equation}
We see that $Z(r,0)$ ranges from values from $10^3$ to $10^7$ within the halo, with the
smallest value $Z_{\rm min}$ attained near the core radius $r_{\rm c}$.  Notice also that
$Z(r)$ reaches even larger values near the origin during the second inner infall exhibited
in the lower left panel of Figs.~\ref{fig:rho} and~\ref{fig:uavg}.

\begin{figure}[ht]
  \begin{center}
    \includegraphics[width=10cm]{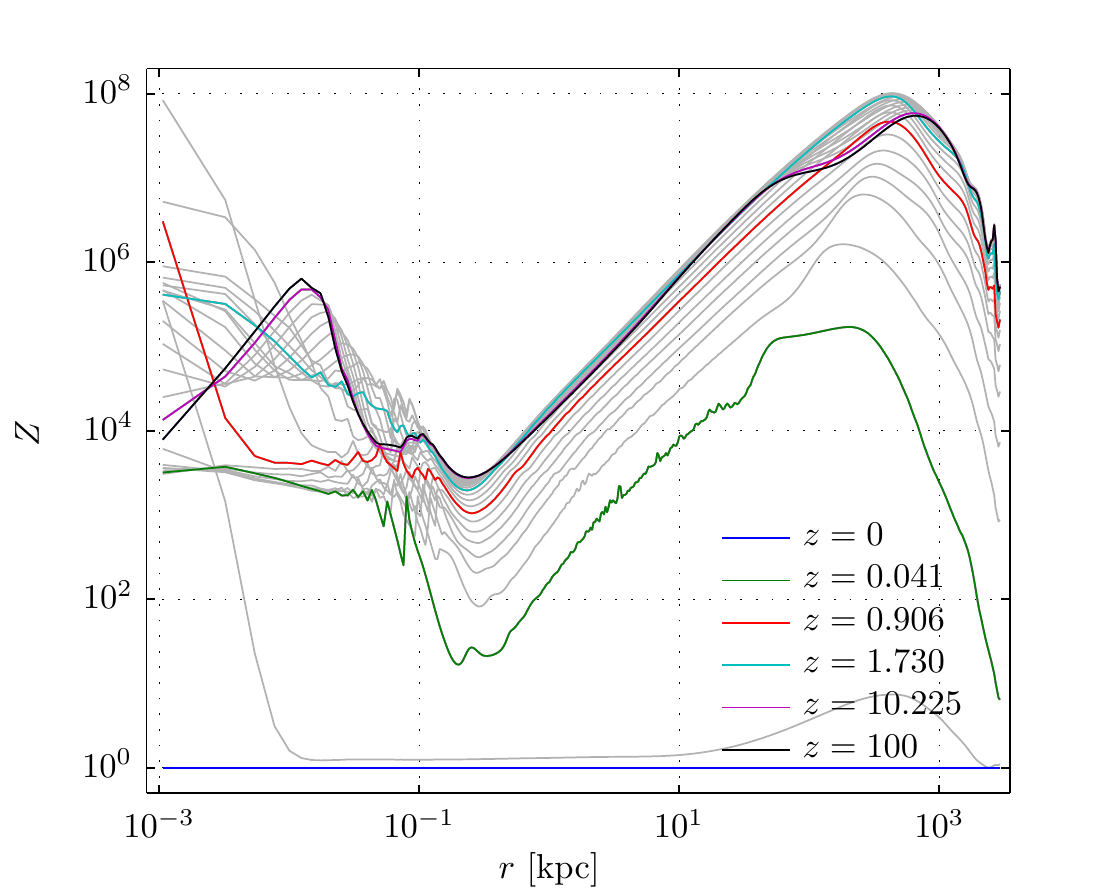}
    \caption{Evolution of the local relaxation factor $Z(r)$, the ratio of the initial value of
      the (pseudo) phase--space density to its value at redshift $z$. Gray lines correspond to
      intermediate redshifts.}
    \label{fig:Z}
  \end{center}
\end{figure}

\subsection{Non--virialized cores}\label{sec:W}

An observer at rest with the collapse center assigns to the DM fluid the kinetic energy
density (recall that $\rho$ is the comoving mass density and that
$3\sigma^2=\sigma_r^2+2\sigma_\theta^2$ is the total comoving velocity dispersion)
\begin{equation*}
  \begin{split}
    {\cal K} &= \frac12 a^{-3}\rho\,\avg{|\bv_{\rm phys}|^2} = \frac12 a^{-3}\!\!\int d^3v\,
    |\,a H\br + \bv/a|^2 f(\br,\bv,s)\\ &= \frac12 a^{-5}\rho\left(a^4H^2r^2 + 
      a^2H r \avg{u} + \avg{u}^2 + 3\sigma^2 \right)
  \end{split} 
\end{equation*}
Recalling the definitions of $\Phi_{\rm M}$ and $\Phi_\Lambda$, the gravitational
potentials generated by the matter background and the cosmological constant, respectively [see
Eq.~(\ref{eq:phi})], the same observer assigns to the fluid the potential energy density
\begin{equation*}
  {\cal U} = - a^{-5}\rho\, r\partial_r\left(a^2\Phi_{\rm M} + \phi\right) + 
  a^{-3}\rho\,\Phi_\Lambda = -\frac12 a^{-5}\rho \left(a^4H^2r^2 + r\phi'\right)   
\end{equation*}
where in the last step we used Friedmann equation (\ref{eq:aF}). To be precise, ${\cal
  U}(r)$ is not a real density, since, through the gravitational potential $\Phi_{\rm M}$,
it depends on all values $\rho(r')$ for $r'\le r$.  But for our purposes, the important
property of ${\cal U}(r)$ is that it does not depend on $\rho(r')$ for $r'> r$, thanks to
Gauss' law.

In the uniform Universe, at redshifts large enough that matter fluctuations can be
neglected but WDM is already non--relativistic, the two energy densities reduce to
\begin{equation*}
  {\cal U} = - \frac{\rhoM}{2a}\,H^2 r^2 \;,\quad
  {\cal K} = - {\cal U} + \frac{3\rhoM}{2 a^5}  \sigma_0^2
\end{equation*}
and ${\cal K}+{\cal U}=0$ if $\sigma_0=0$, as in CDM, consistently with the assumed open
and flat Universe. Anyway, a nonzero $\sigma_0$ does not spoils flatness, since it
contributes a negligible fraction of the total kinetic energy at large distances. However,
the kinetic energy density due to the velocity dispersion dominates at short distances,
since the Hubble flow and the gravitational self--interaction become negligible as
$r\to 0$. In other words, in the central region of the future collapse the kinetic energy
initially overwhelms the potential energy as the $r-$dependent initial virial ratio
\begin{equation*}
  W(r) = \frac{2 K(r)}{-U(r)} = \frac{2 \int_0^r dr'\,r'^2{\cal K}(r')}{- 
  \int_0^r dr'\,r'^2\,{\cal U}(r')} = 2 + \frac{10\sigma_0^2}{a^4H^2r^2}
\end{equation*}
diverges as $r\to0$. This fact is only slightly perturbed by the initial matter
fluctuations, such as those at $z=z_{\rm i}=100$, the initial redshift of our setup. In the
subsequent gravitational collapse triggered by those fluctuations, $W(r)$ changes instead
dramatically.

\begin{figure}[ht]
  \begin{center}
    \includegraphics[width=15cm]{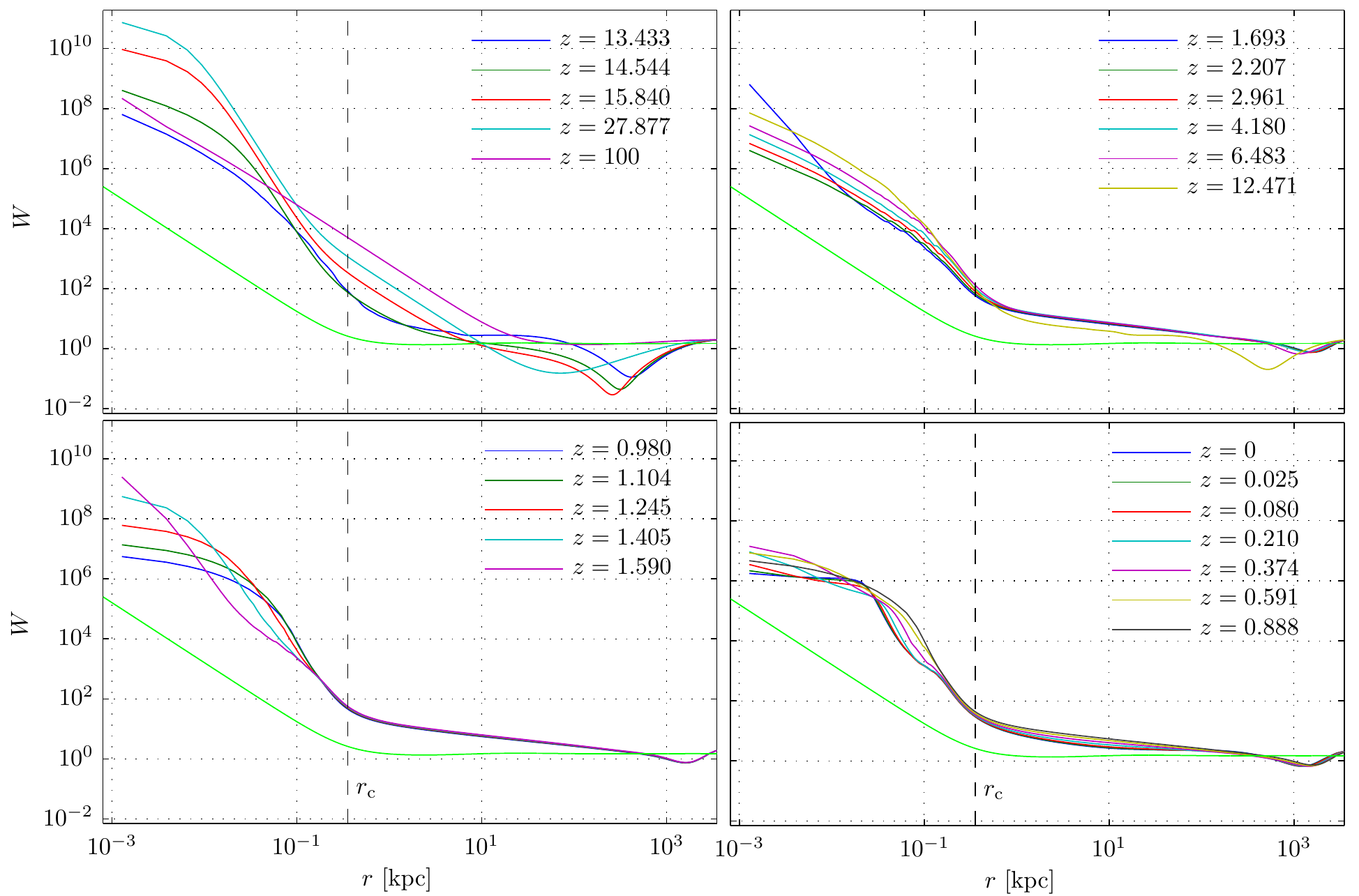}
    \caption{Evolution of local virial ratio $W(r)$ in H1, the same halo of
      Fig.~\ref{fig:rho} and Figs.~\ref{fig:uavg}--\ref{fig:sigt}. The bright green line
      represents the virial ratio of an isothermal sphere with the same core size.}
    \label{fig:W}
  \end{center}
\end{figure}

\medskip 

The common lore on the evolution of $W(r)$, mostly based on the radial collapse model of
CDM, envisage a violent relaxation by which the halo is formed as a gravitationally bound
system, while most matter that is too energetic is lost. The halo is thus left in a
quasi--stationary state that is to a large degree virialized, in the sense that $W(r)\sim
1$ for $r_{\rm c} < r < r_{\rm vir}$, where $r_{\rm c}$ is the core radius and $r_{\rm
  vir}$ is the virial radius, beyond which matter is typically not in a quasi--stationary
state and which could be used to define the halo border. Thus the system should be more
equilibrated if not even thermalized, the closer to the origin one gets. At small
distances, the phase--space distribution function should then tend to an ergodic form
which depends only of the one--particle energy corresponding to the quasi--stationary
self--consistent potential.  The halo density $\rho(r)$ is then expected to monotonically
decrease away from the origin. In this respect, angular momentum has received attention
mostly as a source of corrections to the radial infall model (see {\em e.g.} \cite{zb} and
references therein).

\medskip 

Our numerical results clearly put forward quite a different scenario for spherically
symmetric WDM collapses. The collisionless, non--dissipative VP dynamics, the conservation
of angular momentum and the large values of the initial $W(r)$ (see discussion in the
Sec.~\ref{sec:nonvir}) are in our opinion the theoretical basis of our findings, although
we do not attempt here any quantitative theoretical analysis. The evolution of the virial
ratio $W(r)$, shown in Fig.~\ref{fig:W}, features an excess of kinetic energy at small
distances which gets somehow trapped well inside the halo as the density peak rises at
$r=r_{\rm c}$. $W(r)$ is stabilizing, as expected in a quasi--stationary state, although
with more fluctuations than density or potential. Yet a full plateau of very high $W(r)$
values remains at $z=0$ inside the core, clearly exceeding the power law $(r/r_{\rm c})^2$
proper of equilibrium systems. To remark this fact, we have plotted also the virial ratio
of an isothermal sphere with the same core size of H1. Outside the core $W(r)$ quite
slowly decreases reaching values of order one far away from the core. Thus the halo core
is hollow and really not virialized.

It is conceivable that this quasi--stationary state is only metastable and that the excess
kinetic energy might eventually escape, together with a certain amount of matter and
angular momentum. This certainly does not happen within the finite amount of time
available from the initial redshift $z=100$ to $z=0$. It would be interesting to find the
time scale of the above energy loss, but this is beyond the scope of this work and perhaps
beyond our numerical possibilities, because of the unavoidable diffusion and dissipation
of VP solvers. Moreover, the large tangential motions inside the core of the collapsing
halo and the relatively large value of $Q_{\rm prim}$, as discussed in the Introduction,
suggest that a proper quantum treatment of angular momentum could be necessary to achieve
better quantitative agreement with observations.

As mentioned above, the virial ratio $W(r)$ decreases quite slowly as $r$ grows. If we
define the virial ratio $r_{\rm vir}$ as the distance at which $W(r)$ crosses unity, we
find on average values forty times larger than $r_{100}$. This rather unphysical situation
is due to the excessive extension to large distances of the $1/r^2$ decrease of the
density. The latter is probably due to a bias in our method of selecting the initial
overdensity profiles, leading to an insufficient matter outflow. Improvements in this
respect are certainly necessary, but the hollow core structure can hardly depend on the
lack of a significantly faster density decrease in the halo outskirts. Most likely, a
faster decrease might enhance the core size, rather than reduce it.

\subsection{Hollow cores and initial velocity dispersion}\label{sec:sigma} 

\begin{figure}[ht]
  \begin{center}
    \includegraphics[width=10cm]{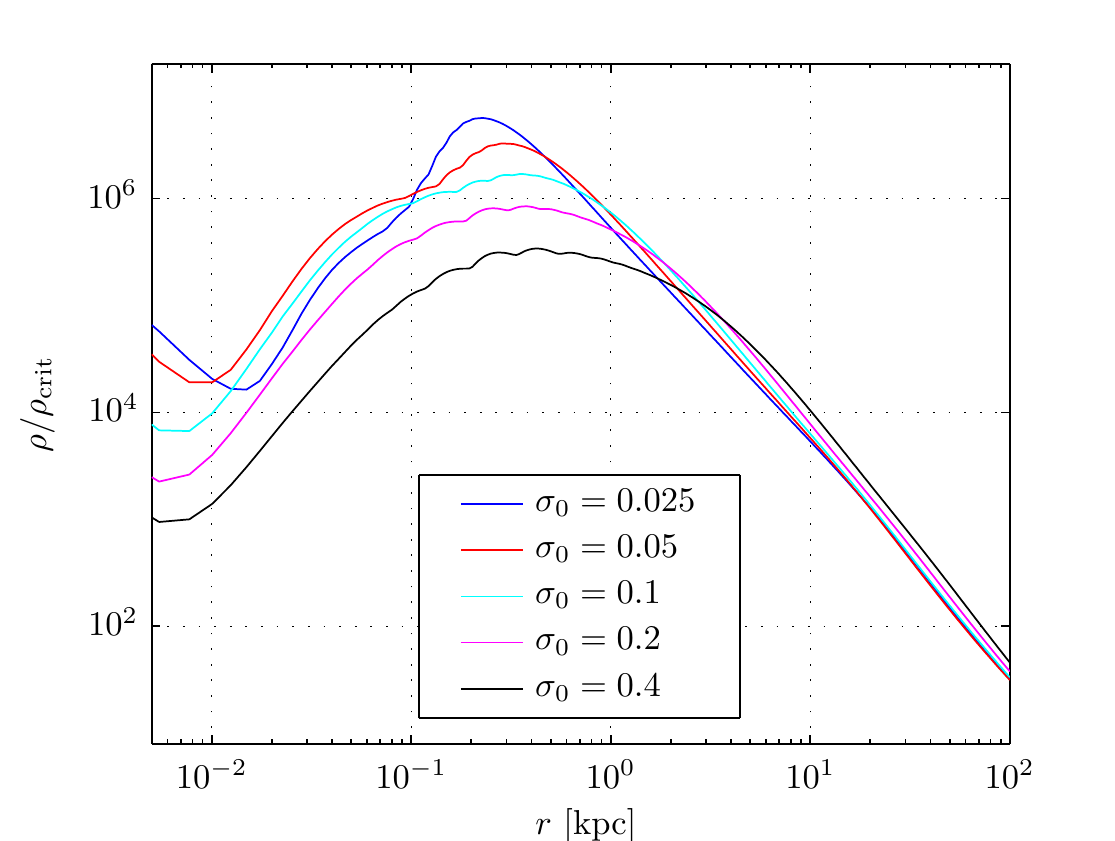}
    \caption{Density curves at redshift $z=0$ for different values of the initial velocity
      dispersion $\sigma_0$, in km/s, but for the same fluctuation profile $\delta_{\rm
        i}(r)$ of halo H1. }
    \label{fig:sigma}
  \end{center}
\end{figure}

It is very interesting to investigate what happens when $\sigma_0$ is varied. In
Fig.~\ref{fig:sigma} we plot the $z=0$ densities for five increasing values of $\sigma_0$,
from $\sigma_0=0.025\,$km/s to $\sigma_0=0.4\,$km/s, obtained always from the same initial
fluctuation $\delta_{\rm i}(r)$ of halo H1. To decrease $\sigma_0$ in our numerical
treatment of the VP equation is definitely more difficult, since smaller $\sigma_0$
implies larger, and very fast increasing, dynamical range in the collapse.  In other
words, more violent relaxation. This can be appreciated from the second column of
Table~\ref{tab:sigma}, which reports the minimal value $Z_{\rm min}$ of the local
relaxation factor $Z(r,0)$ of Eq.~(\ref{eq:Z}).

\bgroup
\def\arraystretch{1.3}%
\begin{table}
\begin{tabular}{|c||c||c|c|c||c|c|c||c|}
  \hline
  $ \sigma_0\,$s/km &$~Z_{\rm min}~$&$~r_{\rm c}/$pc~~&$ \mu_{\rm c}\,{\rm pc}^2/M_\odot$&$ 
  V_{\rm max}\,$s/km&$~r_{\rm 0B}/$pc~~&$ \mu_0\,{\rm pc}^2/M_\odot $&$ 
  V_{\rm B,max}\,$s/km &$ m c^2/{\rm keV} $ \\
  \hline
  $ 0.025 $&$ 2813 $&$ 296  $&$ 203 $&$ 63 $&$ 504 $&$ 603 $&$ 58.2 $&$ 1 $ \\
  \hline
  $ 0.05  $&$ 533 $&$ 413 $&$ 166 $&$ 67.9 $&$ 725 $&$ 517 $&$ 65.9 $&$ 0.594 $ \\
  \hline
  $ 0.1  $&$ 96.8  $&$ 502 $&$ 117 $&$ 73.3 $&$ 1100 $&$ 415 $&$ 72.4 $&$  0.353 $ \\
  \hline
  $ 0.2  $&$ 17.9 $&$ 566 $&$ 68.5 $&$ 78.7 $&$ 1768 $&$ 294 $&$ 77.5 $&$ 0.210 $ \\
  \hline
  $ 0.4 $&$ 4.37 $&$ 598 $&$ 33.6 $&$ 83.6 $&$ 2939 $&$ 188 $&$ 79.8 $&$ 0.125 $ \\
  \hline
\end{tabular}
\caption{Parameter values for halo H1 at different values of the initial velocity
  dispersion $\sigma_0$. In the last column the equivalent WDM mass according 
  to Eq.~(\ref{eq:sigma}).} 
\label{tab:sigma}
\end{table}

Let us first notice that using the $m-\sigma_0$
relation, Eq.~\ref{eq:sigma}, the value $\sigma_0=0.4\,$km/s corresponds to
$m=0.125\,$keV/$c^2$. In turns, this mass corresponds to the $N-$body simulation WDM3
of ref.~\cite{smco5}. One can check that their WDM3 density profile agrees quite well with
our hollow $\rho(r)$ for $r\gtrsim 1\,$pc, on the right of the maximum. This shows
that, on the common domain of validity, $N-$body simulations and direct VP integration
agree, as far as the mass density is concerned.

Fig.~\ref{fig:sigma} and Table~\ref{tab:sigma} show that the overall size of the core
increases with $\sigma_0$, as is natural to expect. Maybe less obvious is the peak
broadening, that progressively makes the notion of hollowness less appropriate and harder
to detect with low--resolution means. In other words, the second length scale of the
hollow core, the hollowness scale, grows to values comparable to the first scale $r_{\rm
  c}$. Notice that $r_{\rm c}$, as defined in Sec~\ref{sec:rc}, does not grow very
much, while $r_{\rm 0B}$, the core radius of the Burkert fit, grows almost linearly with
$m$. Moreover, both $\mu_{\rm c}=\rho_{\rm c}r_{\rm c}$ and $\mu_{\rm c}=\rho_{\rm
  0B}r_{\rm 0B}$ considerably decreases, roughly maintaining their ratio constant.

From the data in Table~\ref{tab:sigma} we can write to a good approximation
\begin{equation*}
  \mu_{\rm c}\frac{{\rm pc}^2}{M_\odot} \simeq  205 - 
  43.6\, y \;,\quad   \mu_0\frac{{\rm pc}^2}{M_\odot} \simeq  614 - 
  105\, y \quad,\quad y = \log_2\left(\frac{40\,\sigma_0}{{\rm km/s}}\right)
  = -\frac43 \log_2\left(\frac{m c^2}{\rm keV}\right)
\end{equation*}
and
\begin{equation}\label{eq:r0sig}
  \frac{r_{\rm 0B}}{\rm pc} = 134 + 348\, \frac{\rm keV}{m c^2} \;.
\end{equation}
Thus, baldly extrapolating to larger values of the WDM mass, we obtain
\begin{equation*}
  \begin{split}
    \mu_{\rm c}=263\,M_\odot/{\rm pc}^2 \;,\quad \mu_0=754\,M_\odot/{\rm pc}^2 \;,\quad 
    r_{\rm 0B} &=355\,{\rm pc} \qquad {\rm when} ~~m=2\,{\rm keV}/c^2\;. \\
    \mu_{\rm c} =308\,M_\odot/{\rm pc}^2 \;,\quad \mu_0=855\,M_\odot/{\rm pc}^2 \;,\quad 
    r_{\rm 0B} &=239\,{\rm pc} \qquad {\rm when} ~~m=3.3\,{\rm keV}/c^2\;.
  \end{split}
\end{equation*}
Since $\mu_{\rm 0,obs}=140^{+83}_{-52} M_\odot$/pc$^2$, we see that the core of H1 would
have a surface density (or surface gravity acceleration, since the two are proportional)
roughly 6 times larger than real DM halos when $m=3.3\,{\rm keV}/c^2$. Taking into account
that this applies to pure WDM collapse (that is with no baryon feedback of any type) and
within a purely classical approach (probably a more relevant limitation), we consider this
a very good starting point for the improvements outlined in the Introduction.

\medskip

The last row in Table~\ref{tab:sigma} shows values of $r_{\rm 0B}$, $\mu_0$ and $V_{\rm
  B,max}$ in very good agreement with observations. The corresponding mass, $m=0.125\,{\rm
  keV}/c^2$, is however too small w.r.t. the constraints from large--scale structure.

As a matter of fact, it is also too small w.r.t. the value $m=1\,{\rm keV}/c^2$ used in
the matter power spectrum to generate the fluctuation field from which the initial profile
was drawn. We cannot pretend to realistically lower the surface density of one given halo
by just raising $\sigma_0$, while keeping fixed the initial overdensity profile. Larger
$\sigma_0$ implies more free--streaming in the linear regime of gravitational clustering,
with the associated small--scale depression of the matter power spectrum. Thus, a
fluctuation profile that was generic at $\sigma_0=0.025\,$km/s, could be practically
impossible at $\sigma_0=0.4\,$km/s. This so--called {\sl catch 22} of WDM, already
mentioned in the Introduction, was pointed out already in ref.~\cite{smco5}, on the basis
of the results of $N-$body simulations. However, in ref.~\cite{smco5}, the scaling of the
core size with the mass was determined to be $r_{\rm core} \sim m^{-2}$, while we find the
relation (\ref{eq:r0sig}) for the Burkert core radius and an even slower decrease for
$r_{\rm c}$. Once extrapolated to $m=1\,{\rm keV}/c^2$, the scaling $r_{\rm core} \sim
m^{-2}$ leads to core sizes around $10\,$pc instead of several hundreds, as we find.

The incorrect scaling $r_{\rm core} \sim m^{-2}$ can be traced to the underestimation of
the relaxation factor $Z(r,0)$, that is the overestimation of the pseudo phase--space
density $Q(r)$. The smallest value $Z_{\rm min}=4.37$ in the second column of table
Table~\ref{tab:sigma}, which corresponds to a value of the mass quite close to simulation
WDM3 of ref.~\cite{smco5}, implies that $Z_{\rm min}$ will be even closer to unity, its
theoretical minimum, for smaller masses. Indeed the WDM $N-$body simulations of
ref.~\cite{smco5,smco6} could resolve the core only for such masses and found $Q$ in the
core very close to its theoretical maximum. But the explosive growth of $Z_{\rm min}$ in
the second column of table Table~\ref{tab:sigma} shows that their extrapolation of $Q$ to
larger masses was ill founded.

\begin{acknowledgments}
The author is deeply indebted to H.J. de Vega and N.G. Sanchez for many illuminating
discussions and for the kind hospitality in the Observatoire de Paris and the Chalonge
School events, were basic motivations were laid down and the development of fruitful ideas
was initiated.

The author thankfully acknowledges the computer resources and technical support provided
by the Physics Department and the INFN section of the University of Milano-Bicocca. 
\end{acknowledgments}

\end{document}